\documentclass[12pt]{article}
\usepackage{color}
\usepackage{graphicx}
\usepackage{amssymb}
\usepackage{breqn}
\usepackage{bbm}
\usepackage{eufrak}
\usepackage[font=small,labelfont=bf]{caption}

\title{\
Towards \textcolor{black}{solving the ICRH wave and \\ Fokker-Planck equations self-consistently}}
\author{Dirk Van Eester, Vincent Maquet, Fabrice Louche \textcolor{black}{and Bernard Reman}\\ 
Laboratorium voor Plasmafysica - Laboratoire de Physique des Plasmas \\
Belgian "EUROfusion Consortium" Member \\
Trilateral Euregio Cluster\\
Renaissancelaan 30 Avenue de la Renaissance\\
B-1000, Brussels, Belgium \\
}
\begin{document}

\maketitle

\abstract{
\textcolor{black}{T}he present paper sketches a 
\textcolor{black}{framework for} solving the wave and Fokker-Planck equations in the ion cyclotron resonance frequency domain fully selfconsistently. 
\textcolor{black}{
It illustrates this can be done by first constructing "building blocks" that are commonly needed by the wave and Fokker-Planck equations, allowing e.g. to account for wave coupling in plasmas containing non-Maxwellian distributions. 
Up to details, 
 the paper} exploits known expressions and methods to solve the 2 intimately connected aspects of the description of the wave-particle interaction underlying ion cyclotron resonance heating.
 Two cases are presented: the case where the guiding centre motion is limited to just following magnetic field lines, and the extended case accounting for drifts away from magnetic surfaces but assuming axisymmetry. A limited set of analytical results is included. As combining wave and Fokker-Planck solving is the focus, \textcolor{black}{the} computation of the dielectric response for arbitrary \textcolor{black}{distribution  functions} is illustrated \textcolor{black}{as well}.}

\section{Introduction}

In magnetic confinement fusion modelling of the wave-particle interaction underlying ion cyclotron resonance heating\textcolor{black}{,} it is standard to separate the time scales to come up with separate equations describing the (fast time scale) electric field pattern associated with a given set of distribution functions of the species the plasma is composed of, and the \textcolor{black}{(slow time scale)} evolution of those distribution functions under the net influence of the radio frequency electric fields. These 2 aspects are intimately connected and should strictly be treated on the same footing. \textcolor{black}{In practice, the assumptions made differ so the obtained solutions of the 2 equations are not fully selfconsistent; most typically it is only the integrated power density on each magnetic surface that is shared.} To properly account for finite temperature and parallel as well as perpendicular gradients effects, it is customary to rely on Fourier transformation of the unknown electric field, find appropriate expressions accounting as realistically as possible for the particle orbits and the local wave-particle interaction, and finally calculate the inverse Fourier transform to get back to physical space.
That inverse transformation quickly becomes tedious, however\textcolor{black}{. Hence,} finding alternative means to solve the actual integrodifferential wave equation is of key importance. Correctly modelling finite Larmor radius effects for modes that break the assumption that the wave phase varies only modestly over a Larmor gyration ($k_\perp \rho_{Larmor} <<1$ where $k_\perp$ is the perpendicular wave number and $\rho_{Larmor}$ is the Larmor radius) is a necessity \textcolor{black}{since that assumption is frequently violated: The very reason for applying ICRH in magnetic confinement fusion machines is to create fast particle populations that \textcolor{black}{do not satisfy} this condition ... And fusion-born populations - again at the very heart of fusion research - typically violate it as well. But rigorously describing the wave-particle interaction quickly leads to } expressions
 containing higher order derivatives - 
 the treatment of which becomes cumbersome when going beyond 1D-application - or even forces to tackle the integrodifferential wave equation "as is". Inspired by the work of Jaeger \cite{AORSA}, some recent efforts omit that last step and simply propose to replace the continuous Fourier-space integral by a sufficiently dense discrete set of modes covering the region of $\vec{k}$-space where modes are admitted. Opposite to what Jaeger did (which is to retain all possible couplings between Fourier modes, yielding a massive linear equation system matrix to be inverted), the computation effort needed can be reduced by only retaining the truly important couplings \cite{DVEfastAORSA}. Rather than a full matrix, the system matrix then (partly) becomes sparse. Although the absolute needs increase, the relative benefit is more and more significant the more spatial dimensions are accounted for. 
\textcolor{black}{
Another classical technique relies on the use of local - rather than global - basis functions: the finite element method.} 
For a suitable \textcolor{black}{grid of points and set of local basis functions}, a development in "quasimodes" \textcolor{black}{defined via these functions was proposed \cite{DVE_Shelf_RF,DVE_quasimodes} to bridge the two.} This method combines the possibilities (i) to write a system in a form completely identical to that obtained when using finite elements (except for the way the coefficients of the equation are assembled), (ii) to account for the intricate $\vec{k}$-space expressions \textcolor{black}{proposed already decades ago to describe the impact of the confining magnetic field on the charged particles' orbits and thereby its effect on the } wave-particle interaction and (iii) allows to cut out weak couplings to speed up the computation. Closer to the more traditional approach is what Bud\'{e} proposed: When it is preferred to study the wave equation solely in physical space, he proposed to solve the integro-differential wave equation as a high order partial differential equation by making a high order polynomial fit of the dielectric response in $\vec{k}$-space so that the inverse Fourier transform becomes trivial (see \cite{Bude,DVEmyBude,DVE_Bude_2D}).    

Opposite to what is done in the present paper - in which well-known classical results are exploited while no new physics is presented \textcolor{black}{and} the burden is maximally put on the computer but analytical development is kept to a strict minimum - there is a lot of recent effort either taking a fresh look at the equations at hand and pushing the analytical treatment further or exploring alternative philosophies. Svid\textcolor{black}{z}inski \cite{Svidzinski} formulated the hot plasma dielectric response in configuration space avoiding approximations by numerically computing the plasma conductivity kernel based on the solution of the linearized Vlasov equation in nonuniform magnetic field. By carrying out needed expressions analytically, Machielsen, Lamalle and Fukuyama proposed a method seeking to speed up this kernel approach while aiming at keeping the physics description as general as possible \cite{Lamalle_kernel_0,Lamalle_kernel,Machielsen,Fukuyama}. In these works an exact solution for the conductivity kernel is derived in configuration space. Machielsen's expressions are valid to all orders in Larmor radius, and up to arbitrary cyclotron harmonic. Lamalle puts the accent on treating the parallel gradients accurately but combined with his earlier work \cite{Lamalle_longpaper} Larmor radius corrections can be included. Practical implementation of the most recent findings with the kernel approach is ongoing \cite{Bernard_kernel}.  Also seeking solutions for the coupled wave and Fokker-Planck equations is not new. Various authors have interfaced wave and Fokker-Planck equation solvers. A few examples are the coupled TORIC and SSFPQL codes by Brambilla and Bilato \cite{TORIC-SSFPQL,Bilato}, the EVE-AQL suite by Dumont \cite{EVE} and the combined AORSA and CQL3D codes by Jaeger and Berry \cite{AORSA_CQL3D,AORSA_CQL3D_2}. An example of the exploitation of the CYRANO wave code \cite{LamalleCYRANO} providing the RF power depositions to solve a set of  coupled Fokker-Planck equations for various beam and wave heated populations is the ICRH package in the EUROfusion European Transport Solver (see e.g. \cite{ETS,ETSICRH}). 

\textcolor{black}{
The combination of the linearised Maxwell Vlasov equations with the quasilinear Fokker-Planck equation naturally and importantly leads to consider non-Maxwellian velocity-space distribution functions. 
The idea of the present paper is to offer a fully selfconsistent approach, based on common building blocks needed in the wave and Fokker-Planck equations.
As a result, part of the focus is on the computation of the dielectric response for arbitrary distribution functions and its practical implementation.}
With the availability of increasingly powerful finite element software \textcolor{black}{(examples are the FEM packages NGsolve \cite{NGsolve} and MFEM \cite{MFEM})}, there are also efforts exploiting such packages. Examples are the \textcolor{black}{works} of Vallejos, Maquet and Zhang \cite{Vallejos,Maquet_NGsolve,Zhang}. 
\textcolor{black}{The present paper offers a self-contained account of existing, important results pertaining to the self-consistent modelling of wave heating (wave propagation and damping + quasilinear diffusion) in the ion cyclotron resonance frequency domain in fusion plasmas. It intends to help pave the way for making use of current and next generation freewares as powerful numerical tools to combine multiscale modelling of plasmas, thereby allowing to concentrate on the underlying physics while exploiting the present-day know-how in computer science. 
}
We underline that the paper does - up to details - not propose any new physics\textcolor{black}{, nor new equations. It merely proposes a scheme allowing to gradually upgrade the physics while guaranteeing self-consistency.}  

The paper is structured as follows: Some basic expressions are reminded in section 2. Section 3 is devoted to discussing the case where drifts away from the magnetic surface are neglected. Section 4 generalises to the case of the axisymmetric tokamak retaining drifts. Section 5 discusses the 2 types of contributions to the basic integrals: It describes the wave-particle interaction on the slow and the fast time scales. Section 6 provides 2 examples of computations of the typical building blocks \textcolor{black}{necessary} to assemble the wave and Fokker-Planck equations. Section 7 is devoted to a brief summary and a minimum of discussion of the road ahead.

\section{Reminder of basic expressions}

We will adopt the notations standardly used when describing wave dynamics in tokamaks so $R$ is the major radius, $Z$ the height above \textcolor{black}{or} below the midplane, $\theta$ is the poloidal and $\varphi$ the toroidal angle. The subscripts "//" and "$\perp$" are with respect to the static magnetic field $\vec{B}_o$. When wave directionality in the plane perpendicular to the magnetic field line is studied the angle $\psi=tan^{-1}[k_{\perp,2}/k_{\perp,1}]$ is introduced where $\vec{k}$ is projected on suitably chosen orthogonal unit vectors $\vec{e}_{\perp,1}$ and $\vec{e}_{\perp,2}$. Finally, the subscript "o" either refers to the static component (in the case of the confining magnetic field), the slowly varying component (in the case of the distribution function) or the magnetic axis (in the case of the major radius), and $\Theta$ is the angle between the parallel and the toroidal direction so $cos \Theta=B_{o,\varphi}/B_o$ and $sin \Theta=B_{o,\theta}/B_o$ so that $\vec{e}_{//}=cos \Theta \vec{e}_\varphi + sin \Theta \vec{e}_{\theta}$ with 
$\vec{e}_\varphi=\partial \vec{x} /\partial \varphi /|\partial \vec{x} /\partial \varphi|$ and 
$\vec{e}_\theta=\partial \vec{x} /\partial \theta /|\partial \vec{x} /\partial \theta|$. $\rho$ is the magnetic surface labeling parameter (half the width of the magnetic surface in the midplane is a frequent choice) so $\nabla \rho / |\nabla \rho|$ is a unit vector perpendicular to a magnetic surface. \textcolor{black}{Furthermore}, $q$ is the charge, $m$ the mass, $N$ the cyclotron harmonic and $n$ the toroidal mode number. We adopt the symbol $\mathcal{N}$ for the density.

\textcolor{black}{Describing the wave-particle interaction involves physical ($\vec{x}$) as well as velocity ($\vec{v}$) space. The wave equation is formulated in physical space so to obtain the wave equation coefficients an integration over velocity is required first.} The Fokker-Planck equation is formulated in terms of constants of the motion while averaging out the 3 types of rapid oscillations (gyromotion, bounce motion and toroidal precession). For an axiymmetric tokamak the 3 most commonly used constants of motion are the energy $\epsilon=mv^2/2$, the magnetic moment $\mu=mv_\perp^2/[2B_o]$ and the toroidal angular momentum (see e.g. \cite{Balescu}) 
\begin{equation}
p_\varphi=\Psi-2\pi m R v_\varphi /q=\Psi-2\pi m R_o v_{//} B_{\textcolor{black}{o,\varphi}} /[qB_o]
\end{equation}
in which $\Psi(\rho)$ is the poloidal magnetic flux, which is sometimes used as flux surface labeling parameter instead of the minor radius $\rho$. 

\textcolor{black}{The absorbed power as defined in the Fokker-Planck equation and that in the wave equation must be identical. This requires the adopted models to be sufficiently sophisticated and handled using the same approximations.} 
The RF absorbed power for a specific specie can be written as
$$P_{abs,FP}=\frac{\partial }{\partial t} \Bigg [ \int d\vec{v}d\vec{x} \epsilon F_0\Bigg ] \Bigg |_{RF} =   \int d\vec{v}d\vec{x} \epsilon \frac{\partial F_o}{\partial t}_{RF}  =  \int d\vec{v}d\vec{x} \epsilon Q(F_o)  = $$
$$\frac{1}{2} Re\Bigg [  \textcolor{black}{\int} d\vec{v}d\vec{x} \epsilon \nabla_{\vec{v}} . \frac{q}{m}\Big [ \vec{E}+\vec{v}\times\vec{B}]^* \int_{-\infty}^{t}dt' \frac{q}{m}\Big [ \vec{E}+\vec{v}\times\vec{B}\Big ] .\nabla_{\vec{v}}F_o\Bigg ]= $$
$$- \frac{1}{2} Re\Bigg [  \textcolor{black}{\int} d\vec{v}d\vec{x} [ \nabla_{\vec{v}}\epsilon]  .\frac{q}{m}\Big [ \vec{E}+\vec{v}\times\vec{B}]^* \int_{-\infty}^{t}dt' \frac{q}{m}\Big [ \vec{E}+\vec{v}\times\vec{B}\Big ] .\nabla_{\vec{v}}F_o\Bigg ]= $$
$$ \frac{1}{2} Re\Bigg [  \textcolor{black}{\int} d\vec{v}d\vec{x} [ \nabla_{\vec{v}}\epsilon]  .\frac{q}{m}\Big [ \vec{E}+\vec{v}\times\vec{B}]^* f_{RF}\Bigg ]=  \frac{q}{2} Re\Bigg [  \textcolor{black}{\int}d\vec{v}d\vec{x}\vec{v} .\Big [ \vec{E}+\vec{v}\times\vec{B}]^* f_{RF}\Bigg ]=$$
\begin{equation} 
- \frac{q}{2} Re\Bigg [  \textcolor{black}{\int}d\vec{v}d\vec{x}\vec{v} .\vec{E}^* f_{RF}\Bigg ]=  - \frac{1}{2} Re\Bigg [  \textcolor{black}{\int}d\vec{x}\vec{J}_{plasma}^* .\vec{E} \Bigg ]=P_{abs,WE}
\label{Eq.Pabs}
\end{equation}
in which the perturbed distribution function can be written in terms of the slowly varying distribution function $F_o$ as
\begin{equation} 
f_{RF}=-\int_{-\infty}^{t}dt' \frac{q}{m}\Big [ \vec{E}+\vec{v}\times\vec{B}\Big ] .\nabla_{\vec{v}}F_o
\end{equation}
while the quasilinear RF diffusion operator is
\begin{equation} 
Q(F_o)=\nabla_{\vec{v}} . \frac{1}{2} Re\Bigg [  \frac{q}{m}\Big [ \vec{E}+\vec{v}\times\vec{B}]^* \int_{-\infty}^{t}dt' \frac{q}{m}\Big [ \vec{E}+\vec{v}\times\vec{B}\Big ] .\nabla_{\vec{v}} \Bigg ] F_o  
\end{equation}
\textcolor{black}{A fully selfconsistent description of the wave and Fokker-Planck equations becomes possible when first evaluating common elementary "building blocks" representing the interaction between an orbit and the wave. }

Rewriting the expressions by referring to the guiding center rather than the particle position allows to account for the particle motion in an elegant way. It is also customary to move to $\vec{k}$-space, which brings in the Kennel-Engelman operator $L_N$ \cite{KennelEngelmann}. Assuming the generator drives oscillations at frequency $\omega$, the 2 relevant Maxwell equations are
\begin{equation} 
\nabla \times \vec{E}=-\frac{\partial}{\partial t} \vec{B}=i\omega \vec{B}  
\end{equation}
$$\nabla \times \vec{B}=\mu_0(\vec{J}_{plasma}+\vec{J}_{antenna})=$$
\begin{equation} 
\mu_0\frac{\partial}{\partial t} \vec{D}+\mu_0\vec{J}_{antenna}=-i\omega \mu_0 \epsilon_0 \overline{\overline{K}}.\vec{E}+\mu_0\vec{J}_{antenna}
\end{equation}
so that 
\begin{equation} 
\nabla \times \nabla \times \vec{E}=i\omega \nabla \times \vec{B}=k_0^2 \overline{\overline{K}}.\vec{E}+i\omega \mu_0\vec{J}_{antenna}
\end{equation}
\textcolor{black}{in which $ \overline{\overline{K}}$ is the dielectric tensor.} In weak variational \textcolor{black}{form} a test function vector $\vec{F}$ is introduced and an integration on the domain is performed. 
The curl-curl term $\vec{F}.\nabla \times \nabla \times \vec{E}$ is substituted for $\nabla \times \vec{E}.\nabla \times \vec{F}^*-\nabla . [\vec{F}^*\times \nabla \times \vec{E}]$ in which the former is symmetric in $\vec{E}$ and $\vec{F}$ and the latter yields the divergence of the Poynting flux when substituting the test function by the electric field so that the variational form of the wave equation becomes the corresponding power balance. The explicit expression for the plasma response \textcolor{black}{ - as will be shown later in the paper - is equally symmetric in $\vec{E}$ and $\vec{F}$, with differential operators acting on both. Hence it does not provide the actual dielectric tensor since }that  would require an expression where no operator acts on the test function so that a proper operator solely acting on the electric field can be defined. \textcolor{black}{The underlying difference in philosophy is crucial: To ensure a fully self-consistent description of the wave particle interaction as treated by the wave and Fokker-Planck equations, the natural reference position is that of the guiding centre and not that of the particle. For the wave equation to directly become the power balance that is common with the expression needed in the Fokker-Planck equation also the test function vector $\vec{F}$ appearing in the variational formulation of the wave equation needs to be evaluated at the guiding centre position so that the needed expression for the dielectric response is an operator acting both \textcolor{black}{on} $\vec{F}$ and $\vec{E}$ and not on $\vec{E}$ alone. This is not imperative, though: Provided the necessary surface/flux terms are included, an actual dielectric tensor can be written such that the wave equation can be written in non-variational form (see e.g. \cite{TOMCAT}, although the actual solving there is done in variational form as well). But doing so hides the symmetry that becomes apparent when treating the 2 equations on fully equal footing.} To get the actual dielectric tensor\textcolor{black}{,} the Kennel-Engelmann operator needs to be removed from the test function, which will give rise to a supplementary flux: the kinetic flux describing the flux carried by particles in coherent motion with the wave.

It is customary to solve both the wave and Fokker-Planck equations leaning on variational techniques. The inspiration comes  from the above expression for the absorbed power. We generalise \textcolor{black}{the second line in Eq.\ref{Eq.Pabs} to}
\begin{equation} 
\mathcal{P}=\frac{1}{2} Re\Bigg [  \textcolor{black}{\int} d\vec{v}d\vec{x}  \textcolor{black}{ G \nabla_{\vec{v}}} . \frac{q}{m}\Big [ \vec{F}+\vec{v}\times \frac{\nabla \times \vec{F}}{i\omega}]^* \int_{-\infty}^{t}dt' \frac{q}{m}\Big [ \vec{E}+\vec{v}\times \frac{\nabla \times \vec{E}}{i\omega}\Big ] .\nabla_{\vec{v}}F_o\Bigg ]
\end{equation}
in which $G$ 
is the test function for the Fokker-Planck equation \textcolor{black}{(i.e. $F_o$ will be written in terms of base functions living in the same space as the test function $G$)} and $\vec{F}$ is the test function vector for the wave equation \textcolor{black}{(i.e. $\vec{E}$ will be written in terms of base functions living in the same space as the test function vector $\vec{F}$)}. Ensuring $G$ is sufficiently continuous, a partial integration allows to get an expression in which $G$ and $\vec{F}$ appear completely symmetrically to $F_o$ and $\vec{E}$:
\begin{equation} 
\mathcal{P}=-\frac{1}{2} Re\Bigg [ \textcolor{black}{\int} d\vec{v}d\vec{x} \textcolor{black}{\textcolor{black}{ [\nabla_{\vec{v}}G]} . \Bigg (  \frac{q}{m}\Big [ \vec{F}+\vec{v}\times \frac{\nabla \times \vec{F}}{i\omega}]^*} \int_{-\infty}^{t}dt' \textcolor{black}{\frac{q}{m}\Big [ \vec{E}+\vec{v}\times \frac{\nabla \times \vec{E}}{i\omega}\Big ] .\nabla_{\vec{v}}F_o} \Bigg ) \Bigg ]
\end{equation}
\textcolor{black}{Lacking sufficient continuity of the adopted local base functions when solving the Fokker-Planck equation, this expression needs to be supplemented with the surface term appearing when performing this partial integration.} To arrive at practical expressions the nabla operator in velocity space is expressed in terms of a chosen set of variables, and the Lorentz force term is written in terms \textcolor{black}{of} guiding centre variables to more easily be able to separate the fast and slow time scales. Both the writing of the required Lorentz force expression \textcolor{black}{(in which one typically uses Faraday's law to eliminate $\vec{B}$ in favour of $\vec{E}$)} in terms of guiding centre variables and dotting the result with the gradient operator have been done by many authors (see e.g \cite{Stix,Swanson,Karney,Ichimaru}), the first expressions dating back to decades ago. 
The easiest way to proceed is to first write the electric field in terms of its $\vec{k}$-spectrum. Strictly this needs to be done using the same reference frame i.e. adopting Cartesian $\vec{k}$ components. In practice one most often adopts a suitable local frame, hereby implicitly assuming that curvature effects are of secondary importance i.e. ignoring the fact that the unit vectors on which the equations are projected are position-dependent so that a proper treatment involves corrections involving metric tensor components and their derivatives (i.e. Christoffel symbols); more often than not integrals along orbits are performed in an approximate way by "patching" local or quasilocal solutions together i.e. using homogeneous or quasi-homogeneous plasma expressions. Since there is no divergence term in physical space we can fake it and adopt a 6-dimensional $\nabla$ operator by padding the factors multiplying $\nabla_{\vec{x}}$ with zero's. Then we can transform the wave-plasma interaction term in variational form to any desired set of variables in the 6-dimensional $(\vec{x},\vec{v})$-space. The general expression for the divergence and gradient operators in $\texttt{N}$-dimensional space is
$$\nabla . \vec{w}=\frac{1}{J_{\textcolor{black}{\vec{\xi}}}} \sum_{j=1}^{\texttt{N}} \frac{\partial}{\partial \xi_j} [J_{\textcolor{black}{\vec{\xi}}} w^j]=\frac{1}{J_{\textcolor{black}{\vec{\xi}}}} \sum_{j=1}^{\texttt{N}} \frac{\partial}{\partial \xi_j} [J_{\textcolor{black}{\vec{\xi}}} \vec{w}.\nabla \xi_j]$$
\begin{equation} 
\nabla H=\sum_{j=1}^{\texttt{N}} \nabla \xi_j \frac{\partial H}{\partial \xi_j}
\end{equation}
in which $J_{\textcolor{black}{\vec{\xi}}}$ is the Jacobian of the transformation and $\xi_i$ are the \textcolor{black}{\texttt{N}} chosen variables.  So labelling the Lorentz \textcolor{black}{acceleration-type} vector as $\vec{\mathcal{H}}(\vec{w})=q/m \Big [ \vec{w}+\vec{v}\times \nabla \times \vec{w}/[i\omega]\Big ] $ the most general expression for $\mathcal{P}$ is
$$\mathcal{P}=\int d\vec{\xi} J_{\textcolor{black}{\vec{\xi}}} \textcolor{black}{G} \nabla_{\textcolor{black}{\vec{\xi}}} . \Big [   \vec{\mathcal{H}}^*(\vec{F}) \int_{-\infty}^{t}dt'  \vec{\mathcal{H}}(\vec{E}) . \nabla_{\textcolor{black}{\vec{\xi}}} \Big ]  F_o=$$
 \begin{equation} 
\int d\vec{\xi} J_{\textcolor{black}{\vec{\xi}}} \sum_{j,j'=1}^6
 \frac{\partial \textcolor{black}{G}}{\partial \xi_{j}}  \nabla_{\textcolor{black}{\vec{\xi}}} \xi_{j} .   \vec{\mathcal{H}}^*(\vec{F}) \int_{-\infty}^{t}dt'  \vec{\mathcal{H}}(\vec{E}). \nabla_{\textcolor{black}{\vec{\xi}}} \xi_{j'} \frac{\partial}{\partial \xi_{j'}}  F_o  
\end{equation}
in which $\vec{\xi}$ is a set of 6 suitable independent variables and $J_{\textcolor{black}{\vec{\xi}}}$ is the corresponding Jacobian\textcolor{black}{, and where the continuity of the flux and the base functions has been assumed implicitly. \textcolor{black}{Surface term contributions need to be added when adopting base functions that do not guarantee this continuity.}} Note that while there is a Jacobian in front as well as behind the partial derivative inside the integral in the equations in non-variational form, there is only a single Jacobian in the variational form. Note as well that the same quantity $\vec{\mathcal{H}}(\vec{E}). \nabla \xi_{j}$ appears in front of as well as inside the \textcolor{black}{orbit} integral. In an axisymmetric tokamak there are 3 constants of motion $\vec{\Lambda}$ (on the RF time scale) and 3 periodic aspects that can be parametrized by a suitable angle $\vec{\Phi}$. The Fokker-Planck equation is formulated in terms of the former 3 and hence requires averaging over the latter 3. In an axisymmetric tokamak the individual toroidal modes $n$ are independent (uncoupled) and can be treated one by one; similarly the cyclotron harmonics $N$ are independent. The poloidal coupling cannot be neglected and requires a bounce average.

Anticipating the generalisation that is discussed later, the next step is to express the RF Lorentz force in terms of its components projected on the gradients of the chosen independent variables $\vec{\mathcal{H}}(\vec{E}). \nabla \xi_{j}$. 
After casting the expression in $\vec{k}$-space \textcolor{black}{and exploiting Faraday's law to eliminate the RF magnetic field}, the relevant Lorentz RF acceleration term is traditionally written in the form $\vec{\mathcal{H}}(\vec{w}_{\vec{k}})=q/[m\omega] \Big [\vec{w}_{\vec{k}}[\omega-\vec{v}.\vec{k}] + [\vec{v}.\vec{w}_{\vec{k}}]\vec{k} \Big ] $. 
The component in the direction of $\nabla_{\vec{v}}\phi=\vec{e}_\phi/v_\perp \partial /\partial \phi $ drops out inside the integral since $F_o$ does not depend on it. The term in front of the time integral also drops out but for a different reason: Just like for the independence of the toroidal modes $n$ in an axisymmetric machine, the double sum on the various test function and electric field cyclotron harmonics reduces to a single sum since the integral of $exp[i(N-N')\phi]$ ($N$ of the electric field and $N'$ of the complex conjugate of the test function vector) is only finite when $N=N'$\textcolor{black}{.}

\section{Traditional treatment: neglecting drifts}

\subsection{Choice of variables }

To stay as close as possible to standard development at the outset, we will start by omitting drift effect\textcolor{black}{s.}
 This has the benefit that only well known expressions appear and that all algebra is straightforward and has already been done by many authors in the past \textcolor{black}{(see e.g \cite{Stix,Swanson,Karney,Ichimaru})}. Generalisation to include drift effects will be discussed later in the text. From first principles the relevant integral is of the form
\begin{equation} 
\int d\vec{\xi}\sum_{j,j'=1}^6
 J \frac{\partial G}{\partial \textcolor{black}{\xi}_{j}}  \Big [\nabla \xi_{j} .  \vec{\mathcal{H}}^*(\vec{F}) \int_{-\infty}^{t}dt'  \vec{\mathcal{H}}(\vec{E}). \nabla \xi_{j'} \frac{\partial}{\partial \xi_{j'}} \Big ] F_o  
\end{equation}
so, using 
\begin{equation} 
|\nabla \Psi | =| \frac{d\Psi}{d\rho}| |  \nabla \rho|  =| \frac{d\Psi}{d\rho}| |  \frac{\partial \vec{x}}{\partial \theta} \times \frac{\partial \vec{x}}{\partial \varphi} /J_{\rho,\theta,\varphi}| = 
| \frac{d\Psi}{d\rho}| |  \frac{\partial \vec{x}}{\partial \theta} \times R \vec{e}_{\varphi}/J_{\rho,\theta,\varphi}| 
\end{equation}
we get 
\begin{equation} 
|B_{o,\theta}|=\frac{|\nabla \Psi|}{2\pi R} 
\end{equation}
\textcolor{black}{It is customary to make a number of transformations to suitable variables. A first transformation is in physical space, going from Cartesian to cylindrical variables $(R,Z,\varphi)$; $R$ is the major radius, $Z$ is the position w.r.t the midplane and $\varphi$ is the toroidal angle; the associated \textcolor{black}{$\vec{x}$-space} Jacobian is simply $R$. Next, we go to a frame accounting for the magnetic geometry, introducing the poloidal magnetic flux $\Psi$ as magnetic surface labeling parameter (the poloidal magnetic flux is only a function of the more widely used minor radius i.e. $\Psi(\rho)$) and the poloidal angle $\theta$.
Similarly, the Jacobian to go from local adequately rotated "Cartesian" $\vec{v}=(v_{\perp,1},v_{\perp,2},v_{//})$ to "cylindrical" $(v_\perp,v_{//},\phi)$ is $v_\perp$. Then further transforming from $(v_\perp,v_{//},\phi)$ to $(\epsilon,\mu,\phi)$ we use 
$$v_\perp=[2B_o\mu/m]^{1/2}$$
\begin{equation} 
v_{//}=sign(v_{//})[2(\epsilon-\mu B_o)/m]^{1/2}
\end{equation}
to find the intermediate velocity space Jacobian $B_o/[v_{//}v_\perp m^2]$ so that the total velocity space Jacobian from $(v_{\perp,1},v_{\perp,2},v_{//})$ to $(\epsilon,\mu,\phi)$ is $J=B_o/v_{//}m^2$. As $v_{//}=v_\theta B_o/B_{o,\theta}$ we can rewrite this as
\begin{equation} 
J=\frac{B_\theta}{[v_\theta m^2]}=\frac{1}{2\pi m^2 \dot{\theta}J_{\Psi,\theta,\varphi}}
\end{equation}
in which $J_{\Psi,\theta,\varphi}$ is the Jacobian in physical space when using $(\Psi,\theta,\varphi)$ as coordinates, and the relation $|\nabla \Psi|=|\partial \vec{x}/\partial \theta \times \partial \vec{x}/\partial \varphi|/J_{\Psi,\theta,\varphi}=|\partial \vec{x}/\partial \theta|R/J_{\Psi,\theta,\varphi}$ has been exploited. Hence we can write the total coordinate transformation as
\begin{equation} 
d\vec{x} d\vec{v}=d\Psi d\theta d\varphi d\epsilon d\mu d\phi \frac{1}{2\pi m^2 \dot{\theta}}
\end{equation}
}
\textcolor{black}{Introducing the bounce angle that varies linearly with time $\Phi_b=\omega_b t$ (with $\omega_b$ the bounce frequency) we can write the final Jacobian  elegantly and conveniently
\begin{equation} 
J_{\Psi,\theta,\varphi,\epsilon,\mu,\phi}= \frac{1}{2\pi m^2 \omega_b}
\label{JacobianCOM}
\end{equation}
Note that this Jacobian is a constant of the motion, which has the advantage that it can be moved freely across the bounce integral. In absence of drifts corrections, $\Psi$ is a constant of the motion.}

\subsection{Concrete expressions and elementary building blocks }

Many authors have derived practical expressions for the wave-particle interaction. In this section, we will follow Ichimaru's formulation \cite{Ichimaru}. A quantity $\vec{H}$ can be written in terms of its Fourier spectrum
\begin{equation} 
\vec{H}(\vec{x},t)=\int d\vec{k} exp[i(\vec{k}.\vec{x}-\omega t)] \vec{H}_{\vec{k}}
\end{equation}
where $\vec{x}$ is the particle position. Ichimaru wrote a suitable expression for the dielectric response, evaluated in terms of the local slowly varying distribution function and evaluated at the guiding centre position $\vec{x}_{GC}$:
$$\vec{F}^*.\overline{\overline{K}}.\vec{E}=[1-\sum_\alpha \frac{\omega^2_\alpha}{\omega^2}]\vec{F}^*_{\vec{k}'}.\overline{\overline{1}}.\vec{E}_{\vec{k}} -2\pi \sum_\alpha \frac{\omega^2_\alpha}{\omega^2} \sum_{N=-\infty}^{+\infty}  \int_{-\infty}^{+\infty}d\vec{k}'  \int_{-\infty}^{+\infty}d\vec{k} $$
\begin{equation}
 \int_0^{+\infty} dv_\perp \int_{-\infty}^{+\infty}dv_{//} \frac{N\Omega_\alpha \partial F_{o,\alpha}/\partial v_\perp+k_{//}v_\perp \partial F_{o,\alpha}/\partial v_{//}}{N\Omega +\tilde{k}_{//}v_{//}-\omega}L(\vec{F}^*_{\vec{k}'})L(\vec{E}_{\vec{k}})
 \label{Eq1}
 \end{equation}
\textcolor{black}{in which $\alpha$ refers to the type of species.} In the above $\tilde{k}_{//}$ either is the $k_{//}$ of the electric field or - inspired on more complete theory \cite{Kaufman1972} and worked out by Lamalle when isolating the dominant contributions \cite{Lamalle_longpaper} - the mean values of the $k_{//}$'s of the test function and the electric field. Moreover, 
\begin{equation} 
L_N(\vec{H})=v_\perp \Big [ H_-J_{N+1}exp[i(N+1)\psi]+H_+J_{N-1}exp[i(N-1)\psi] \Big ]+v_{//}H_{//}J_Nexp[iN\psi]
\end{equation}
with $E_{\pm}=E_{\perp,1}\pm iE_{\perp,2}$\textcolor{black}{. The subscripts "1" and "2" underline the fact that the wave equation is written in a suitable set of field-aligned components with 1 ($//$) component along the magnetic field and 2 in the plane perpendicular ($\perp$) to it. Following Jaeger \cite{AORSA}, an elegant choice is to opt for 2 perpendicular directions, one of which is as close as possible to the unit vector in the major radius direction $\vec{e}_R$ so that the second is close to $\vec{e}_Z$}.  Writing the phase in terms of the guiding centre generalises the phase to $\Phi=\vec{k}.\vec{x}_{GC}(t)-N \phi_{gyro}(t)-\omega $.  The guiding centre motion is written as $d\vec{x}_{GC}/dt=\vec{v}_{GC}$ where $\vec{v}_{GC}$ is the guiding centre drift velocity. The Larmor gyration is captured through $d\phi_{gyro}/dt=-\Omega$. Adopting the zero-banana width limit the expression for the velocity reduces to the simpler $d\vec{x}_{GC,//}/dt=\vec{v}_{//}$. Omitting the fact that $\vec{e}_{//}$ is actually not a constant vector, this locally further reduces to the elementary $dx_{GC,//}/dt=v_{//}$; upgrades are possible but require careful accounting of the changes of the adopted unit vectors on which the equations are projected. We will adopt that simplified expression so that the relevant phase can be written as $\Phi=k_{//}x_{//}-N\phi-\omega$. 
The corresponding quasilinear diffusion operator can be rewritten as (see e.g. \cite{KennelEngelmann,Stix, Swanson, Karney}) 
\textcolor{blue}{
}
\begin{equation} 
Q(F_o)=\int d\vec{k}\int d\vec{k}' \pi q^2 m^2 \mathcal{L}_N \Bigg [  \delta(\omega-\tilde{k}_{//}v_{//}-N\Omega)  L^*_N(F_{\vec{k}'})L_N(E_{\vec{k}})  \mathcal{L}_N(F_o) \Bigg ] 
\label{Qdef}
\end{equation}
in which the differential operator $\mathcal{L}_N$ is
\textcolor{blue}{
}
\begin{equation} 
\mathcal{L}_N=\frac{1}{\omega m^2v_\perp}[(\omega-\tilde{k}_{//}v_{//}) \frac{\partial}{ \partial v_\perp} +\tilde{k}_{//} v_\perp\frac{\partial}{\partial v_{//}}]
=\frac{1}{\omega m^2v_\perp}[N\Omega\frac{\partial}{ \partial v_\perp} +\tilde{k}_{//} v_\perp\frac{\partial}{\partial v_{//}}]
\label{LNdefinition}
\end{equation}
and where the equality stems from the fact that there is only a resonant contribution. Up to the factor $\omega m^2 v_\perp$ in the denominator this operator is the one appearing in Ichimaru's expressions for the dielectric response; we will label the latter as $\tilde{\mathcal{L}}_{N}$. After writing the \textcolor{black}{expression given in Eq.\ref{Qdef}} in variational form and after a partial integration, the RF term in the Fokker-Planck equation is 
\begin{equation} 
- \int d\vec{x}d\vec{v}
\int d\vec{k}\int d\vec{k}' \pi q^2 m^2 \mathcal{L}_N(G)  \delta(\omega-\tilde{k}_{//}v_{//}-N\Omega)  L^*_N(F_{\vec{k}'})L_N(E_{\vec{k}})  \mathcal{L}_N(F_o) 
\end{equation}
which is symmetric in the test function $G$ and the unknown distribution $F_o$, because the original quasilinear term already shows this symmetry.

We will show the full correspondence between the expressions relevant for the Fokker-Planck and the wave equation. The basic equations are
$$P_{abs,WE}=\int d\vec{x}\frac{1}{2}Re[\vec{E}^*.\vec{J}]=\int d\vec{x}\frac{1}{2}Re [\frac{k_o^2\vec{E}^*.\overline{\overline{K}}.\vec{E}}{i\omega \mu_o}]=\int d\vec{x}\frac{1}{2}Re [-i\omega\epsilon_o\vec{E}^*.\overline{\overline{K}}.\vec{E}]$$
\begin{equation} 
P_{abs,FP}=\int d\vec{x}d\vec{v} \frac{mv^2}{2}Q(F_o)
\end{equation}
We start by evaluating the absorbed power when having the quasilinear operator $Q(F_o)$ \textcolor{black}{(Eq.\ref{Qdef})}. Multiplying with $\epsilon=mv^2/2$ and integrating over velocity space yields - after a partial integration and since $\mathcal{L}_N(mv^2/2)=1/m$ - we get
$$P_{abs,FP}=\int d\vec{v} \frac{mv^2}{2} Q=$$
\begin{equation} 
 -\frac{1}{2} Re \Bigg [   \int dv_\perp dv_{//} \frac{2\pi}{m\omega} \Big [ \pi q^2  \delta(\omega-k_{//}v_{//}-N\Omega)  L^*_N(E_{\vec{k}'})L_N(E_{\vec{k}})  \tilde{\mathcal{L}}_N(F_o) \Big ] \Bigg ] 
\end{equation}
The expression we adopted for the dielectric response \textcolor{black}{is an operator that originates from substituting the test function vector $\vec{F}$ by the electric field $\vec{E}$.
 As that operator acts both on test function and on the field (as opposed to only acting on the electric field, allowing to write the equation in non-variational form), it does not directly correspond to $\vec{E}^*.\vec{J}$ 
where the perturbed current density $\vec{J}$ is linked to the dielectric tensor 
$\overline{\overline{K}}$ via 
$\vec{J}=-i \omega \epsilon_o \overline{\overline{K}} . \vec{E}$. } 
Correcting for the needed factor \textcolor{black}{$-i\omega \epsilon_o$ while using Eq.\ref{Eq1}} we get
$$P_{abs,WE}=\frac{1}{2}Re \Bigg [ \frac{\omega \epsilon_o}{i}(-2\pi) \frac{\omega_p^2}{\omega^2}\int dv_\perp \int dv_{//} 
[N\Omega \frac{\partial F_o}{\partial v_\perp}
+k_{//}v_\perp \frac{\partial F_o}{\partial v_{//}} ]
\frac{L^*_N(\vec{E}_{\vec{k}'}) L_N(\vec{E}_{\vec{k}})}{[N\Omega + k_{//}v_{//}-\omega]}\Bigg ] 
$$
\begin{equation} 
=-\frac{1}{2}Re \Bigg [ \int dv_\perp \int dv_{//}  (2\pi) \frac{\mathcal{N}q^2}{ m \omega}
\frac{L^*_N(\vec{E}_{\vec{k}'}) L_N(\vec{E}_{\vec{k}})}{i[N\Omega + k_{//}v_{//}-\omega]} \tilde{\mathcal{L}}_N(F_o)\Bigg ] 
\end{equation}
which is identical to the Fokker-Planck expression since $\omega_p^2=\mathcal{N}q^2/[\epsilon_om]$ \textcolor{black}{in which $\mathcal{N}$ is the density}, the distribution is normalised to $1$ in Ichimaru's expressions and the delta function appears with a factor $\pi$ from encircling the pole.

Looking at the expressions for the absorbed power for the 2 equations, one concludes that  to obtain a fully selfconsistent picture one needs to evaluate the common "building blocks"
\begin{equation} 
 \int d\vec{k} d\vec{k}' \omega_p^2
\frac{\tilde{\mathcal{L}}_N(G) L^*_N(\vec{F}_{\vec{k}'}) L_N(\vec{E}_{\vec{k}})\tilde{\mathcal{L}}_N(F_o) }{i[N\Omega + k_{//}v_{//}-\omega]} 
\end{equation}
Replacing $G$ by the energy for the wave equation, and $\vec{F}$ by $\vec{E}$ for the Fokker-Planck equation one obtains the needed expression describing the wave-particle interaction. 

\section{Upgraded treatment: including drifts}

\subsection{Choice of variables including drifts}

The step to retaining orbit drift effects while assuming axisymmetry requires accounting for the difference between the poloidal flux $\Psi$ and the toroidal angular momentum $p_{\varphi}$. This involves generalising the guiding centre motion from just having a parallel direction component to including the full drift velocity $\vec{v}_{GC}$ and accounting for the dependence of $p_{\varphi}$ on $v_{//}$. The former yields the replacement of $k_{//}v_{//}$ by the more general  $\vec{k}.\vec{v}_{GC}$ in the resonant denominator while the latter requires to upgrade the operator $\tilde{\mathcal{L}}_N$. Balescu derived the explicit expression for the drift velocity \cite{Balescu} and Lamalle computed the expressions for the gradients $\nabla \epsilon$, $\nabla \mu$ and - in particular - $\nabla p_\varphi$ \cite{Lamalle_longpaper} needed to find the more general expression. Following Balescu, the explicit expression of the drift velocity is
\begin{equation} 
\vec{v}_{GC}=v_{//}\tilde{\vec{b}}
\label{vGC}
\end{equation}
with
\begin{equation} 
\tilde{\vec{b}}=\vec{e}_{//}+ \frac{\vec{e}_{//}}{\Omega} \times  \Big [ v_{//} (\vec{e}_{//}.\nabla )\vec{e}_{//}  +\frac{1}{mv_{//}} [\mu \nabla B_o]\Big ] 
\label{veceb}
\end{equation}
where the extra terms - responsible for deviation from magnetic surfaces - constitute the curvature and grad B drifts, respectively, but where the contribution involving the electrostatic potential has been omitted to remain as close as possible to the earlier mentioned result. To be consistent, the time derivative of the gyrophase also has an extra contribution, equally provided by Balescu and involving derivatives of the unit vectors in the 2 adopted perpendicular directions. 

The orbits can be determined \textcolor{black}{by} solving the actual guiding center equation of motion $d\vec{x}_{GC}/dt=\vec{v}_{GC}$. \textcolor{black}{An elegant alternative way to} determine the guiding center orbits is directly from the definition of $p_\varphi$ \cite{DVE_finiteorbitwidth}, however\textcolor{black}{:} For a given energy and magnetic moment, plotting the contours lines of $p_\varphi$ for a prescribed sign of $v_{//}$ produces - by definition - the corresponding orbits.
 Along the orbits $\delta p_\varphi=0$ so they require $\delta R$ and  $\delta Z$ to be related to ensure $p_\varphi$ is conserved: $0=\delta R \partial p_\varphi/\textcolor{black}{\partial}R + \delta Z \partial p_\varphi/\textcolor{black}{\partial}Z $. Hence the orbits are found solving either of the 2 following differential equations: 

\begin{equation} 
\frac{\textcolor{black}{d} Z}{\textcolor{black}{d} R}=- \frac{\partial p_\varphi /\partial R}{\partial p_\varphi/\partial Z}  
\end{equation}
or

\begin{equation} 
\frac{\textcolor{black}{d} R}{\textcolor{black}{d} Z}=- \frac{\partial p_\varphi /\partial Z}{\partial p_\varphi/\partial R} 
\end{equation}
The choice among these 2 depends on which variable is most slowly varying: at maxima or minima of $R$ or $Z$ one of the derivatives diverges while the other tends to zero.  

The upgraded divergence and gradient terms in combined $(\vec{x},\vec{v})$-space are
$$\nabla_6=\nabla_{\vec{x}_{GC}}
+\nabla \mu \frac{\partial }{\partial \mu}
+\nabla \epsilon \frac{\partial }{\partial \epsilon}
+\nabla p_{\varphi} \frac{\partial }{\partial p_{\varphi}}$$   
\begin{equation} 
\nabla_6.\vec{w}=\frac{1}{J_6} \Bigg [ \frac{\partial }{\partial \vec{x}_{GC}} \Big [ J_6 \nabla_{\vec{x}} \vec{x}_{GC}.\vec{w} \Big] 
+  \frac{\partial }{\partial \mu} \Big [ J_6 \nabla \mu .\vec{w} \Big ]
+ \frac{\partial }{\partial \epsilon}  \Big [ J_6 \nabla \epsilon .\vec{w} \Big ]
+\frac{\partial }{\partial p_{\varphi}}  \Big [ J_6\nabla p_{\varphi} .\vec{w} \Big ] \Bigg ] 
\end{equation}
in which the Jacobian is still a constant of the motion as the earlier found velocity space Jacobian $B_o/[v_{//}m^2]$ now generalises to $\tilde{B}_o/[v_{//}m^2]$ where $\tilde{B}_o=B_o \Big [1+[v_{//}/\Omega]\vec{e}_{//}.(\nabla \times \vec{e}_{//}) \Big ]$, a result found by Lamalle \textcolor{black}{\cite{Lamalle_longpaper}} except that he uses $v$ and a normalised $\mu$ rather than $\epsilon$ and $\mu$ adopted more commonly, e.g. by Balescu. The results found by Balescu and Lamalle confirm the earlier derived \textcolor{black}{Eq.\ref{JacobianCOM}}
\begin{equation} 
d\vec{x}d\vec{v}=d\varphi d\phi d\Phi_b d\epsilon d\mu 
dp_{\varphi} \frac{1}{2\pi m^2 \omega_b}
\end{equation}
except that this result now also holds when drift corrections are included, rigorously accounting for the distinction between $\vec{x}$ and $\vec{x}_{GC}$, and between $p_\varphi$ and $\Psi$. 

\subsection{Concrete expressions and elementary building blocks}

The expression showing the full equivalence of the expression for the absorbed power in the wave and Fokker-Planck equations provided at the start \textcolor{black}{can} now be written in an explicit elegant form \cite{DVE_actionangle}
$$\frac{\partial}{\partial t}[\int d\vec{v} \epsilon F_o]|_{RF}=
$$
$$
\frac{(2\pi)^3}{2}Re \Bigg [ \int d\vec{\Lambda} \epsilon \frac{\partial}{\partial \vec{\Lambda}} . \Big [ J \sum_{\vec{\texttt{m}}}  \frac{\vec{h}_{\vec{\texttt{m}}}}{\omega}< \dot{\epsilon}^*_{\vec{\texttt{m}}}e^{-i(\vec{\texttt{m}}.\vec{\Phi}-\omega t)}\int_{-\infty }^tdt' \dot{\epsilon}_{\vec{\texttt{m}}} e^{+i(\vec{\texttt{m}}.\vec{\Phi}-\omega t)} >\frac{\vec{h}_{\vec{\texttt{m}}}}{\omega} .  \frac{\partial F_o}{\partial \vec{\Lambda}} \Big ] \Bigg ]= $$
$$-\frac{(2\pi)^3}{2}Re \Bigg [ \int d\vec{\Lambda} J  \Big [  \sum_{\vec{\texttt{m}}}  < \dot{\epsilon}^*_{\vec{\texttt{m}}}e^{-i(\vec{\texttt{m}}.\vec{\Phi}-\omega t)}\int_{-\infty }^tdt' \dot{\epsilon}_{\vec{\texttt{m}}} e^{+i(\vec{\texttt{m}}.\vec{\Phi}-\omega t)} >\frac{\vec{h}_{\vec{\texttt{m}}}}{\omega} .  \frac{\partial F_o}{\partial \vec{\Lambda}} \Big ] \Bigg ] =$$
\begin{equation} 
\frac{1}{2}Re \Big [ \int d\vec{x} \vec{E}^*.\vec{J}_{RF}\Big ]
\end{equation}
in which the \textcolor{black}{actions $\vec{\Lambda}=(\epsilon/\omega,\tilde{p}_\varphi,-m\mu/q)$ and the associated angles $\vec{\Phi}$} were adopted, $\dot{\epsilon}=d\epsilon/dt=q\vec{E}.\vec{v}$ (to be evaluated along the orbit and hence containing the subtleties of the various dependencies as captured by the Kennel-Engelmann operator) and $\vec{h}_{\vec{\texttt{m}}}=(1,n,-N)$, where $n$ is the toroidal mode number and $N$ the cyclotron harmonic. \textcolor{black}{Although the actual action-angle variables were adopted in that earlier work, there is no need to exploit these; the set closer to the more standard variables can still be exploited. For example, the tilde above the toroidal angular momentum $p_\varphi$ refers to the fact that the relevant quantity derived from first principles has an extra factor $-2\pi/q$ in its definition: $\tilde{p}_\varphi=-2\pi p_\varphi /q$. This extra factor is commonly omitted since it allows the direct identification of the coordinate $\Psi $ and $p_\varphi$ when drifts are neglected.} The non-resonant term $1-\sum_\alpha \omega_p^2/\omega^2$ separated by Ichimaru 
appears in a natural way when exploiting the expression
\begin{equation} 
\frac{q}{m}[\vec{E}+\vec{v}\times \vec{B}]=\frac{i}{\omega m}[\frac{d}{dt}\nabla_{\vec{v}} - \nabla_{\vec{x}}] q\vec{E}.\vec{v}
\end{equation}
for the RF-induced Lorentz force due to Landau \& Lifschitz \cite{Landau}: It can readily be shown that $\int d\vec{v} \dot{\epsilon}\sum_j\nabla_{\vec{v}}\dot{\epsilon}.\nabla_{\vec{v}}\Lambda_j \partial F_o/\partial \Lambda_j=-\mathcal{N}q^2|\vec{E}|^2$; recall that $\omega_p^2=\mathcal{N}q^2/[\epsilon_om]$. 
\textcolor{black}{Separating this nonresonant term associated with the total derivative in the above allowed Ichimaru to obtain expressions only requiring 
$\dot{\epsilon}$, 
the quantity expressed in terms of guiding centre variables by Kennel \& Engelmann.} 
The expression of the part of the perturbed distribution function that gives rise to the resonant denominator 
is
\begin{equation} 
f_{RF,res}=-\sum_{\vec{\texttt{m}}}\int_{-\infty}^tdt'   \dot{\epsilon}_{\vec{\texttt{m}}}  e^{+i(\vec{\texttt{m}}.\vec{\Phi}-\omega t)} \frac{\vec{h}_{\vec{\texttt{m}}}}{\omega} .  \frac{\partial F_o}{\partial \vec{\Lambda}} 
\end{equation}
and the RF diffusion operator is

\begin{equation} 
Q=\frac{1}{J}\frac{\partial}{\partial \vec{\Lambda}} . \Bigg [ J \sum_{\vec{\texttt{m}}}  \frac{\vec{h}_{\vec{\texttt{m}}}}{\omega}< \dot{\epsilon}^*_{\vec{\texttt{m}}}e^{-i(\vec{\texttt{m}}.\vec{\Phi}-\omega t)}\int_{-\infty }^tdt' \dot{\epsilon}_{\vec{\texttt{m}}} e^{+i(\vec{\texttt{m}}.\vec{\Phi}-\omega t)} >\frac{\vec{h}_{\vec{\texttt{m}}}}{\omega} .  \frac{\partial F_o}{\partial \vec{\Lambda}} \Bigg ]  
\end{equation}
A key element to get the above equality is that the partial derivatives of the toroidal and gyro-angles are identical to those using the original $\phi$ and $\varphi$ since the bounce motion correction in an axisymmetric tokamak only involves the bounce angle $\Phi_b$ and the constants of the motion: $\phi=\Phi_g+P(\vec{\Lambda},\Phi_b)$ and $\varphi=\Phi_d+T(\vec{\Lambda},\Phi_b)$ i.e. $\phi$ and $\varphi$ are the "reference" values at the end time $t$ of the orbit integral w.r.t. which the bounce motion is described. \textcolor{black}{In the above the Jacobian is $J=\omega/[m^3\omega_b]$ rather than the earlier obtained result in view of the slightly different variables; the subscripts "g" and "d" refer to the gyromotion and to the toroidal precession motion, respectively.}
The minus sign in $\Lambda_3$ makes that the associated mode number $m_g$ is $-N$ rather than $N$. 

\textcolor{black}{The proposed procedure to arrive at a fully self-consistent description thus is composed of the following steps: (i) Use the expression of Landau \& Lifshitz to write the RF Lorentz force in terms of a non-resonant term 
and a term that involves a gradient of the work $q\vec{E}.\vec{v}$ the electric field does on the particle. (ii) Use the Kennel-Engelman operator to write the latter in terms of this quantity evaluated at the guiding centre position. (iii) Use the action-angle formalism to write the time derivative and the gradient operator in its simplest possible form \cite{DVE_actionangle}. Along the orbit the former is 
$$d/dt=\partial/\partial t +\vec{v}.\nabla_{\vec{x}}+\vec{a}.\nabla_{\vec{v}}=\partial / \partial t + d\vec{\Lambda}/dt.\partial/\partial \vec{\Lambda}+ d\vec{\Phi}/dt.\partial/\partial \vec{\Phi}.$$
Without the RF contribution we simply get $d/dt=\partial / \partial t  +\vec{\omega}.\partial/\partial \vec{\Phi}$ while including it requires
$$d\vec{\Lambda}/dt=d[...]/dt+\sum_{\vec{m}}\frac{d\epsilon_{\vec{m}}}{dt} exp[ i\vec{m}.\vec{\Phi}] \vec{h}_{\vec{m}}/\omega$$
where $\vec{h}_{\vec{m}}=(1,m_d,m_b)$ and the total time derivative term $d[...]/dt$ yields the non-resonant contribution isolated by Ichimaru. 
} 

Although the general treatment of the wave equation in terms of action-angle variables has already been proposed decades ago by Kaufman \cite{Kaufman1972}, its practical realisation is still awaited. \textcolor{black}{O}btaining the Fourier components of the work $\dot{\epsilon}_{\vec{\texttt{m}}}=q\vec{E}.\vec{v}$ the electric field does on the particles requires the evaluation of a very large number of bounce Fourier amplitudes\textcolor{black}{, the evaluation of which is time consuming and therefore commonly avoided. Indeed, as} the variation of $\Omega$ is captured in terms of bounce modes, the number of bounce modes to be retained when describing cyclotron heating is of order $\Omega/\omega_b$, which is typically of order $10^4$ for fusion relevant parameters. The bounce and cyclotron time scales being so far apart, the obtention of $\dot{\epsilon}_{\vec{\texttt{m}}}$ along the orbit requires performing an accurate bounce integral, tracking the variation \textcolor{black}{of} the phase $\vec{k}.\vec{x}_{GC}+N\Omega(\vec{x}_{GC})-\omega t$ on the RF time scale as well as on the bounce time scale. Just because the time scales are so far apart the dominant contributions to the integral are picked up near stationary phase points which yields the usual resonant denominator away from turning points, and its upgrade close to them. In case of Landau or TTMP damping ($N=0$), the most rapidly varying factor of the phase (the factor $N\Omega$) is lacking so stationary phase methods are not adequate. In that case a standard Fast Fourier transform becomes the recommended approach to evaluate the integral. \textcolor{black}{Original work in that respect was provided by Puri \cite{Puri}.} The interested reader can find details in \cite{Lamalle_longpaper,DVE1995,DVE1998,DVE2001}. 

Most closely connected to the usual approach are the expressions obtained when assuming that the stationary phase method \textcolor{black}{holds} throughout. The relevant rapidly varying phase in the Fourier integral 
$H_{\vec{\texttt{m}}}=(2\pi)^{-3}\int d\vec{\Phi}exp[-i\vec{\texttt{m}}.\vec{\Phi}] H$ to obtain the Fourier modes in terms of the angles used in the action-angle formalism  being 
$\vec{\texttt{m}}.\vec{\Phi} -[ \vec{k}.\vec{x}_{GC}-N\phi]$ 
(the Fourier modes are determined at a fixed time $t$) and all 3 the action angles varying linearly with time with respective frequencies $\vec{\omega}$ characteristic for the bounce, Larmor gyration and toroidal precession oscillatory motion, 
the stationary phase condition is
\begin{equation} 
\vec{\texttt{m}}.\vec{\omega}=\vec{k}.\vec{v}_{GC}+N\Omega
\end{equation}
The number of bounce modes being very large, the stationary phase points form a dense set along the orbit so that the sum over the bounce modes can - as a first approximation - be replaced by a continuous bounce integral. Strictly, this is just a - luckily often reasonable - approximation, though. For example close to turning points the density of stationary phase points is insufficient to warrant that replacement to be justified. 

Accepting the fact that the description is not fully complete, a first step towards a more general expression is thus \textcolor{black}{to} incorporate the drift terms in the resonant denominator and use the expressions provided in the previous section. Note that the resonant denominator appearing when performing the orbit integral is $i[\vec{\texttt{m}}.\vec{\omega}-\omega]$ ($\vec{\texttt{m}}.\vec{\omega}=\omega$ is a global rather than a local condition on the orbit) which at the stationary phase point is $i[\vec{k}.\vec{v}_{GC}+N\Omega-\omega]$ i.e. is the generalisation of the resonant denominator $i[k_{//}v_{//}+N\Omega-\omega]$ of the case ignoring drifts. 

A somewhat more rigorous treatment consists in adding the supplementary velocity space gradient arising from the  distinction between $\Psi$ and $p_\varphi$. 
A subtlety arises, however: Writing the earlier introduced $\tilde{\mathcal{L}}_N$ operator in terms of the actions, \textcolor{black}{the 2 earlier given expressions provided in Eq.\ref{LNdefinition}} respectively provide
$$\tilde{\mathcal{L}}_N=mv_\perp \Big [ \frac{N\Omega+k_{//}v_{//}}{\omega}\frac{\partial}{\partial \Lambda_1}+N\frac{\partial}{\partial \Lambda_3}\Big ] $$
and
\begin{equation} 
\tilde{\mathcal{L}}_N=mv_\perp \Big [ \frac{\partial}{\partial \Lambda_1}+\frac{\omega-k_{//}v_{//}}{\Omega}\frac{\partial}{\partial \Lambda_3}\Big ] 
\end{equation}
when expressed in function of the actions. These 2 expressions are only identical when the resonance condition $N\Omega +k_{//}v_{//}=\omega$ is satisfied. In the Fokker-Planck equation this is the case but in the wave equation it is not. However, at the stationary phase point $\vec{\texttt{m}}.\vec{\omega}=\vec{k}.\vec{v}_{GC}+N\Omega$ so very close to it the phase varies as it would when inhomogeneity is absent i.e. 
\begin{equation} 
\vec{\texttt{m}}.\vec{\Phi}(t)-\omega t \equiv \vec{\texttt{m}}.\vec{\omega}t  -\omega t=[\vec{k}.\vec{v}_{GC}+N\Omega-\omega ]t
\end{equation}
and hence (i) the resonant denominator is of the form we recognize from uniform plasma theory but (ii) this only holds very locally near the proper stationary phase point. \textcolor{black}{Provided the phase is indeed very rapidly varying, a significant simplification of the expressions is justified and yields expressions which - up to details - resemble those of homogeneous plasma. When - on the contrary - the phase varies slowly, the details of the motion along the orbit need to be accounted for differently. 
}
 
The term arising from the 3rd constant of motion adds an extra term $n\partial /\partial \Lambda_2$ inside the straight bracket in the expression for $\tilde{\mathcal{L}}_N$. Accounting for the difference in notation this finally yields the generalisation
\begin{equation} 
\tilde{\mathcal{L}}_{N} \rightarrow \tilde{\mathcal{L}}_{N} -\frac{2\pi n}{\omega q}\frac{\partial }{\partial p_\varphi}
\end{equation}
and the associated generalisation of $\mathcal{L}_N$. The obtained operator was already identified by Lamalle \textcolor{black}{\cite{Lamalle_longpaper}}, be it expressed in slightly different constants of motion. Similar to what was obtained for the case drifts were omitted, the symmetry of the original Kennel-Engelmann and differential operators yields symmetries of the "building blocks" to be computed. 
\begin{equation} 
 \int d\vec{k} d\vec{k}' \omega_p^2
\frac{\tilde{\mathcal{L}}_N(G) L^*_N(\vec{F}_{\vec{k}'}) L_N(\vec{E}_{\vec{k}})\tilde{\mathcal{L}}_N(F_o) }{i[N\Omega + \vec{k}.\vec{v}_{GC}-\omega]} 
\end{equation}
Through $p_\varphi$, the velocity gradient operator expressed in terms of the constants of motion now also has a component that describes RF induced radial drifts.

The toroidal angular momentum $p_\varphi$ is only a constant of the motion in an axisymmetric machine. When modelling stellarator plasmas or tokamak plasmas accounting for the ripple, the above expressions relying on them thus are no longer rigorous. On the other hand, the choice proposed by Balescu \textcolor{black}{(Eqs.\ref{vGC} and \ref{veceb}) can still be exploited: It only relies on the fact that the energy $\epsilon$ and the magnetic moment $\mu$ are constants of the motion. In that case, the magnetic surface labeling parameter can be used just like in the driftless case and the usual Ichimaru operator can be exploited; the reference $\Psi$ then merely is a label to identify the orbit, not a constant of the motion. But doing so, the slow time scale dynamics can only be correctly captured in its most general form when actually tracking the orbits on a long timescale, replacing the orbit integral by a piecewise sampling and updating the slowly varying distribution accordingly. When doing so the time integral $\int_{-\infty}^t$  introduced when solving Vlasov equation then has to be replaced by a time stepping procedure, requiring the perturbed distribution function to be updated gradually and likewise adapting the Fokker-Planck equation to reflect this upgrade. Which is a daunting task when done fully rigorously. 
In the simpler case of an axisymmetric machine so that orbits close on themselves in a toroidal cut, tracking the longer timescale history almost always allows a simple separation of timescales. This is the topic of the next section. } 

\section{Fast and slow dynamics integral} 
The obtained "building blocks" needed to assemble the coefficients of the wave and Fokker-Planck equations contain terms 
 of the form 
\begin{equation}
exp[-\textcolor{black}{i}\tilde{\Phi}(t)]\int_{-\infty}^{t}dt' exp[i\Phi(t')]
\label{phases}
\end{equation}
to be evaluated. Here $\tilde{\Phi}$ is the phase of the test function and $\Phi$ is that of the electric field. More often than not, one relies on "adiabatic switch-on" of the electric field to evaluate this integral in its simplest possible form: only retaining the end point contribution assuming the particles "lose memory" of past events. In practice, however, the time evolution can be of importance. In an axisymmetric tokamak the orbits are closed poloidally but not toroidally and single magnetic field lines ergodically cover the full magnetic surface when accounting for a large number of transits. Exceptions are the discrete set of "rational" surfaces where the field line upon finishing a poloidal turn has simultaneously equally finished an integer number of toroidal turns. The latter are of special importance since they allow constructive interference of successive encounters of a resonance. When formulating the wave equation it is customary to ignore the reality and importance of rational surfaces. For making MHD studies, in contrast, these surfaces are the ones the focus is on. \textcolor{black}{Strictly, the integrand of the orbit integral depends on time not only via the rapidly varying phase; e.g. $v_\perp$ and $v_{//}$ vary along the orbit. When the phase is rapidly varying, the dominant contributions to the integral are picked up at stationary phase points and the corrections due to the supplementary time variation are of secondary importance. When the phase is slowly varying, the dynamics along whole orbit needs to be accounted for both in the "fast" and "slow" time integral. In the present section, the more traditional assumption that $\Phi$ is rapidly varying will be adopted.}

\subsection{Slow timescale dynamics}

We split the total time evolution in a sum of contributions, one from each poloidal "bounce"
 \begin{equation} 
exp[-\textcolor{black}{i}\tilde{\Phi}(t)]\int_{-\infty}^{t}dt' exp[i\Phi(t')]=\sum_{M=0}^{\infty} exp[-\textcolor{black}{i}\tilde{\Phi}(t)]\int_{t-(M+1)\tau_b}^{t-M\tau_b}dt' exp[i\Phi(t')]
\end{equation} 
 Here, $\tau_b$ is the time required to make a full "bounce" i.e. to return to the same poloidal position. At each successive bounce - and in an axisymmetric machine - the phase changes by the same amount $\Delta \Phi$ for a given orbit. So we can rewrite the above as 
 \begin{equation} 
exp[-\textcolor{black}{i}\tilde{\Phi}(t)]\int_{-\infty}^{t}dt' exp[i\Phi(t')]=
 \Big [ \sum_{M=0}^{\infty} 
 exp[iM \Delta \Phi ] \Big ] exp[-\textcolor{black}{i}\tilde{\Phi}(t)]\int_{t-\tau_b}^{t}dt' exp[i \Phi(t')]
\end{equation}
 If $\Delta \Phi$ is not a multiple of $2\pi$ or a rational part of it (i.e. steering away from "MHD analysis") then the relevant phase at the successive encounters will have taken all possible values between $0$ and $2\pi$ after an infinity of bounces. Hence - although particles exhibit an energy kick at every encounter with the resonance(s) along their path - there is no net interaction between the electric field and the particles in that case. In practice, however, collisions (or any other decorrelation \textcolor{black}{mechanism}) changes the picture. Labeling $\nu$ to be the collision or decorrelation frequency, the phase now is \textcolor{black}{$\vec{k}.\vec{x}_{GC}(t)-N \phi_{gyro}(t)-[\omega+i\nu]t$} so we get the upgraded expression
 \begin{equation} 
exp[-\textcolor{black}{i}\tilde{\Phi}(t)]\int_{-\infty}^{t}dt' exp[i\Phi(t')]=$$
$$
 \Big [ \sum_{M=0}^{\infty} 
 exp[M (i\Delta \Phi -\nu \tau_b)] \Big ] exp[-\textcolor{black}{i}\tilde{\Phi}(t)]\int_{t-\tau_b}^{t}dt' exp[i \Phi(t')]
\end{equation}
\begin{figure}   
 \begin{center}
 \includegraphics[width=3.1in]{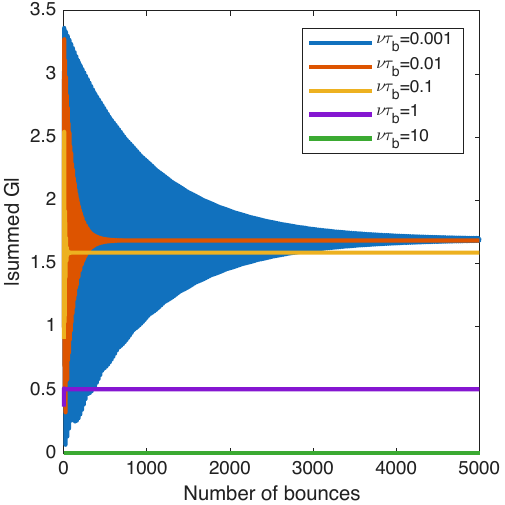}
  \includegraphics[width=3.1in]{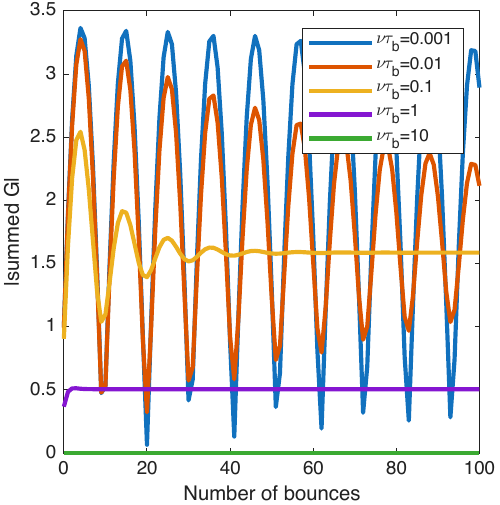}
\caption{Switch-off factor for 6 different values of $\nu \tau_b$; the right plot provides a detail at small $M$.}
 \label{fig:switch-off-factor} 
 \end{center}  
\end{figure}

The switch-off factor $G=\sum_{M=0}^{\infty} exp[M (i\Delta \Phi -\nu \tau_b)]$ was plotted in Fig. \ref{fig:switch-off-factor} for 6 different values of $\nu \tau_b$. For large values, the net effect quickly goes to zero ("superadiabaticity") while for small values the exact value of $\nu$ does not matter after enough bounces have been considered. The latter is very often the case: the bounce frequency and the collision frequency are typically various order of magnitude apart. For intermediate values, some absorption remains but the net effect is diminished. 
\textcolor{black}{Such more detailed treatment relies on full-orbit following }
treatment in which \textcolor{black}{both the details of the slow and fast dynamics matter.}

\subsection{Fast timescale dynamics}

Next we turn to the evaluation of the fast behaviour and look at the remaining integral covering the history from $t'=t-\tau_b$ to $t'=t$. We start by reminding the different time scales appearing in the problem. The above discussion concentrated on the difference between the collisional (or more general decorrelation) timescale and the bounce time scale. The present discussion involves the bounce and the cyclotron timescales. We use typical JET parameters and focus on core $H$ minority fundamental cyclotron heating at $B_o=3.45T$. The thermal velocity is $v_{th}=9.70\times 10^3 [T/A]^{1/2}$ where $T$ is the temperature in $eV$ and $A$ is the atomic mass. Assuming $T=10keV$ that yields  typical velocities of $10^6m/s$. \textcolor{black}{Adopting a circular cross section} for simplicity and concentrating on a magnetic surface $\rho=0.3m$ from the magnetic axis while adopting $sin \Theta=B_{0,\theta}/B_o=0.3$ and $v_{//}/v=0.6$ yields a bounce time of $\tau_b=2\pi \rho/[v_{//}sin \Theta] \approx 10^{-5}s$ while the RF time scale is $\tau_{RF}=1/f\approx 2\times 10^{-8}s$ so about $500$ Larmor gyrations occur during a single bounce motion. Hence, unless a more detailed picture is required the bounce time can be substituted for infinity. 

We will focus on the changes close to the cyclotron resonance. To do that we need to evaluate $\Phi$ more in detail. We make a Taylor series expansion of the phase around a later to be defined reference time $t_o$: $\Phi \approx \Phi_o+ \Phi_1(t-t_o)+ \Phi_2(t-t_o)^2/2$ where
$$ \Phi_1=\dot{\Phi}=k_{//}v_{//}+N\Omega-\omega$$
\begin{equation} 
\Phi_2=\ddot{\Phi}=k_{//}a_{//}+N\dot{\Omega}
\end{equation}
As we are just illustrating the behaviour, the variation of $k_{//}$ has been neglected. We get the needed quantities from the equation of motion and the conservation laws. Since the energy $\epsilon=mv^2/2$ and the magnetic moment $mv_\perp^2/[2B_o]$ are conserved, we have
$$v_{//}=sign(v_{//})[v^2-CB_o]^{1/2}$$
\begin{equation} 
a_{//}=\frac{dv_{//}}{d\theta} \dot{\theta}=-C\frac{\partial B_o}{\partial \theta} \frac{\dot{\theta}}{2v_{//}}
\end{equation}
where $C=v_\perp^2/B_o$ at a reference point and $\dot{\theta}=v_\theta/|\partial \vec{x}/\partial \theta|$; $v_\theta$ is obtained from $v_{\textcolor{black}{\theta}}=sin\Theta v_{//}$ in which $\Theta$ is the angle between the static magnetic field and the toroidal direction $\Theta=cos^{-1}[B_{0,\varphi}/B_o]$. 

We introduce the notations $T=(t'+t)/2-t_o$ and $\tau=t'-t$ and subsequently adopt the specific choice that $t_o$ is the mean time $(t+t')/2$ i.e. that $T=0$ (note that that implicitly assumes the variation of e.g. the magnetic field is mild between $t$ and $t'$). Two types of expressions are adopted in the literature: either one only makes a Taylor series development of $\Phi$ \textcolor{black}{(associated with the phase of the electric field $\vec{E}$)} or - more symmetrically and corresponding to more general theory - of both $\Phi$ and $\tilde{\Phi}$ \textcolor{black}{(the latter being the phase associated with the test function $\vec{F}$)}\textcolor{black}{; see the relevant expression Eq.\ref{phases}}. The former yields an expression in which the electric field's $k_{//}$ appears, the latter causes the mean $\tilde{k}_{//}$ to appear instead. We will denote either choice as $\mathtt{k}_{//}$. After adopting $T=0$ the 2 possible options are
\begin{equation} 
\textcolor{black}{\Phi -\tilde{\Phi} \approx} \Phi_0-\tilde{\Phi}_0+ \frac{\Phi_1}{2}\tau+\frac{\Phi_2}{8}\tau^2 
\end{equation}
 and  
\begin{equation} 
\textcolor{black}{\Phi -\tilde{\Phi} \approx} \Phi_0-\tilde{\Phi}_0+\frac{\Phi_1+\tilde{\Phi}_1}{2}\tau+\frac{\Phi_2-\tilde{\Phi}_2}{8}\tau^2 
\end{equation}
The factors independent of time are the $\vec{k}.\vec{x}$ at the reference position. The factors that are linear in time are the usual factors $N\Omega +\mathtt{k}_{//}v_{//}-\omega$ appearing in the resonant denominator in homogeneous theory. Finally the last factor describes the acceleration or deceleration due to the change of $\Omega$ and $v_{//}$ along the guiding centre orbit. It speaks for itself that the Taylor series expansion only holds for times and positions close to the reference time and position. Significant changes of these parameters require a piece-by-piece "patchwork" evaluation of the time integral when integrating backward in time. We will henceforth adopt the notation $\overline{\Phi}_0+\overline{\Phi}_1\tau+\overline{\Phi}_2\tau^2$ for either of the cases.

We now make several supplementary changes of variables. First we go to $\tau=t'-t$ - the integral then runs from $-\infty$ to $0$ \textcolor{black}{(strictly from $-\tau_b$ to $0$ but recall that since $\Omega >> \omega_b$ one can substitute $\tau_b$ by $\infty$)} - and then we flip the sign of the path variable to get $\tilde{\tilde{t}}=-\tau$ so that the path goes from $0$ to infinity. Then we further transform $\tilde{\tilde{t}}$ to $p=\alpha \tilde{\tilde{t}}+\beta$ so that we can complement the argument of the exponential to a square and write the integrand as $exp[-p^2]$. The value $\alpha \infty$ and the direction of the chosen path ensures the integral converges when approaching times in the further past. Completing the square in the exponent requires
\begin{equation} 
-p^2+\textcolor{black}{\beta^2}=i(-\overline{\Phi_1} \tilde{\tilde{t}}+\overline{\Phi_2} \tilde{\tilde{t}}^2)
\end{equation}
(note the sign change of the linear term) where 
$$ \alpha=e^{-i\pi/4}\overline{\Phi}_2^{1/2}$$   
\begin{equation} 
\beta=i\overline{\Phi}_1/[2 \alpha]
\end{equation}
so that the integral is 
\begin{equation} 
exp \Big [ i[\Phi_0-\tilde{\Phi}_0] \Big ] \frac{\pi^{1/2}}{2\alpha} Erfc(\beta) exp[\beta^2]
\end{equation}
in which $Erfc$ is the complementary error function
\begin{equation} 
Erfc(\beta)=\int_\beta^{\alpha \infty}dp e^{-p^2}
\end{equation}
This smooth function is only varying significantly when the absolute value of its argument is below 2 and it can be considered as a constant elsewhere i.e. the wave-particle response it models is well localised near the resonance where $\Phi_1$ crosses zero. Whatsoever, the adopted Taylor series loses significance too far from a reference point. \textcolor{black}{This e.g. means that details of what the truncated series predicts for times beyond where the Taylor series is a valid approximation have no physical meaning but that a "snap shot" approach for describing the wave-particle interaction by cutting up the time history integral into tiny bits where the series holds makes sense. Consequently, just retaining the first time derivative term in the Taylor series (formally identical to the homogeneous limit) is typically sufficient. Only very close to turning points (where the first time derivative of the phase is zero so that the second time derivative needs to be retained to assess the impact of the inhomogeneity sensed along the orbit) requires an upgraded treatment.} Note that the argument of the function diverges when 
the acceleration term vanishes.
In most cases, this divergence is an artefact without consequence, however: When the quadratic term can justifiably be neglected w.r.t. the linear term (i.e. when $\beta \rightarrow \infty$ so that $Erfc(\beta)=1/[\pi^{1/2} \beta exp(\beta^2)]$) the expression reduces to the familiar homogenous plasma resonant denominator
\begin{equation} 
\frac{1}{i\Phi_1}exp \Big [ i[\Phi_0-\tilde{\Phi}_0]\Big ] =exp[i(\vec{k}'-\vec{k}).\vec{x}]\frac{1}{i[N\Omega + \mathtt{k}_{//}v_{//}-\omega\textcolor{black}{]}}
\end{equation}
The latter expression completely discards past events and merely retains the contribution from the last time $t$ in the path integral. This is the approach most commonly adopted when assembling ICRH wave equations. 

The adopted truncated Taylor series expansion of the phase suggests that the acceleration term $\Phi_2$ is small w.r.t. the velocity term $\Phi_1$ so that it can usually justifiably be dropped. But up to exceptions quite the opposite is true: $\ddot{\overline{\Phi}}=[\partial \dot{\overline{\Phi}}/\partial x_{//}] v_{//}$ so $\overline{\Phi}_1/\Phi_2\approx L_{//}/v_{//}$ where $L_{//}$ is the parallel gradient length, which is typically of order $R$ so typically the acceleration term is several orders of magnitude bigger than the velocity term; for example for a $5keV$ H ion, the thermal velocity is $7\times 10^5m/s$. $L_{//}$ only drops to zero when the guiding centre velocity goes through an extremum: at banana tips and inflection points.  As a consequence, the Taylor series expansion is only a valid approximation when $|\overline{\Phi}_1/\overline{\Phi}_2|>>|\tau|$ which is in a very tiny time interval around the reference position. At the resonance position ($\overline{\Phi}_1=0$) the complementary error function describes the net energy exchange between the electric field and the charged particle while away from it there is an increasingly fast oscillation around a mean value. This has the deep consequence that - up to few exceptions - locally adopting homogeneous plasma expressions suffices to describe the wave-particle interaction (which is extremely convenient as it justifies a "quasi-local" approach in most cases) while any expansion soon loses physical meaning when going away from the reference position (which underlines that the results of more sophisticated models need to be interpreted with care). 

\begin{figure}   
 \begin{center}
 \includegraphics[width=6.1in]{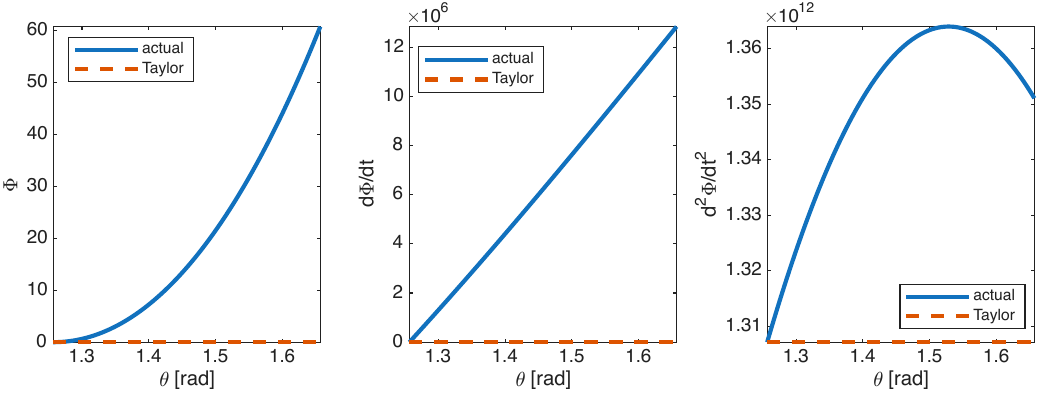}
\caption{\textcolor{black}{Exact phase} $\Phi$ and its first and second time derivatives for a $H$ ion crossing the cyclotron resonance\textcolor{black}{; the dashed line is the phase and its first 2 derivatives evaluated at the reference point (leftmost angle) used to evaluate the Taylor series approximation of  $\Phi$ near the resonance point.}}
   \label{fig:Phi} 
 \end{center}  
\end{figure}

\begin{figure}   
 \begin{center}
 \includegraphics[width=5.1in]{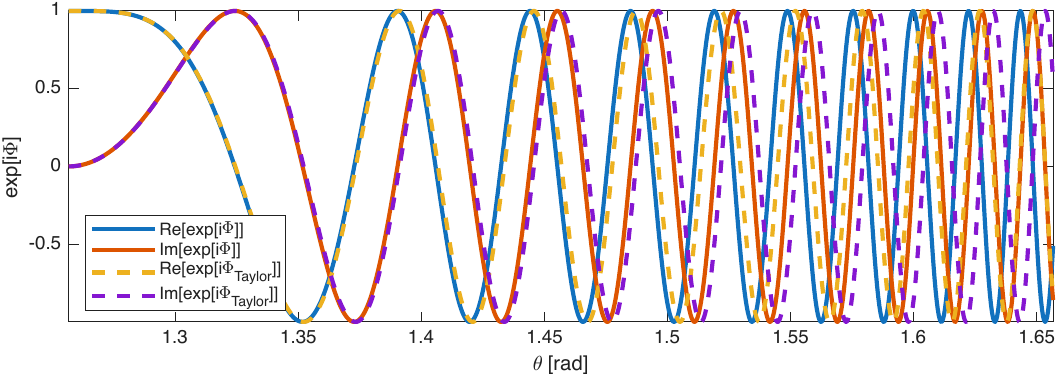}
 \includegraphics[width=5.1in]{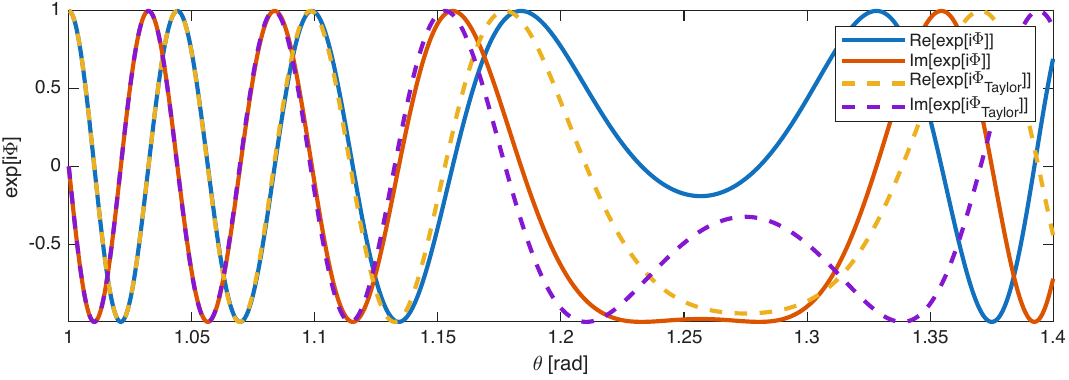}
\caption{Integrand $exp[i\Phi]$ of the "kick" integral at (top) and far from the resonance (bottom).}
  \label{fig:expPhi} 
 \end{center}  
\end{figure}

\begin{figure}   
 \begin{center}
 \includegraphics[width=5.1in]{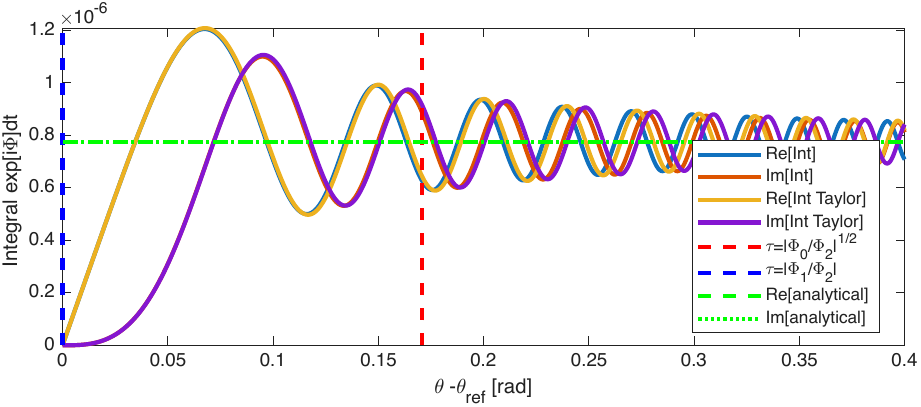}
 \includegraphics[width=5.1in]{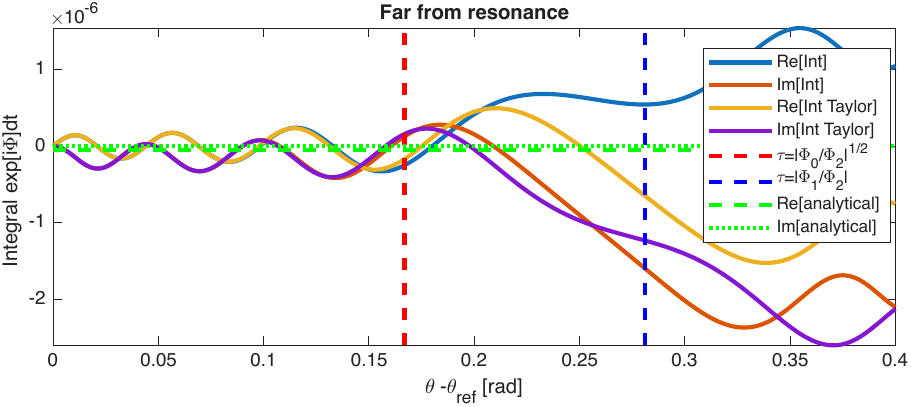}
\caption{Kick integral $\int dt exp[i\Phi]$.}
 \label{fig:intexpPhi2} 
 \end{center} 
\end{figure}

Figure \ref{fig:Phi} depicts the phase for a passing $H$ particle with velocity $v=10^6m/s$ and $v_{//}/v=0.6$ at the equatorial plane in a JET plasma with $f=51MHz$, $B_o=3.45T$, $B_{0,\theta}/B_o=0.3$ adopting a circular cross section. The adopted magnetic surface minor radius is $\rho=0.3m$; JET's major radius is $R_o=2.97m$. For the chosen parameters the duration of an oscillation period at the driver frequency is about $2\times 10^{-8}s$; in that time the poloidal distance covered \textcolor{black}{is} about $10^{-2}rad$. \textcolor{black}{As illustrated by the sharp contrast between the full and dashed lines in Fig.\ref{fig:Phi} the coefficients of the Taylor series expansion change very quickly when deviating from the reference position.} Near the resonance the \textcolor{black}{oscillation slows} down, giving rise to a "kick" when crossing the resonance. It can be seen in Fig.\ref{fig:expPhi} - depicting the evolution of the integrand $exp[i\Phi]$ in a poloidal angle interval of $0.4rad$ - that the approximation using the Taylor series expansion gradually dephases when the position is too far from the reference position (taken at the leftmost angle in the plot). Close to the reference point the Taylor series expansion provides a proper representation but the difference between the exponential factor with the actual or the approximated phase soon becomes visible. Close to the resonance the result is a mild dephasing but far from the resonance the behaviour is completely different. 
In order for the Taylor series expansion to be sufficiently accurate the contribution involving $\Phi_2$ should be small w.r.t. the lower order contributions. In Fig. \textcolor{black}{\ref{fig:intexpPhi2}}  the integral $\int dt exp[i\Phi ]$ is plotted, both when using the actual phase and the approximated phase. The red and blue vertical lines respectively represent the locations at which the quadratic term in $\Phi$ equal the zero order term (i.e. $\tau=|\Phi_0/\Phi_2|^{1/2}$) and at which the quadratic term equals the first order term (i.e. $\tau=|\Phi_1/\Phi_2|$) in the Taylor series expansion. In order for the Taylor series approximation to be acceptably accurate one should only use $\tau$'s (and the corresponding poloidal angles $\theta=\int_0^\tau dt \dot{\theta}$) well smaller than these limiting values. The full lines only represent the integral in a limited interval (here $|\Delta \theta|<0.4rad$). Performing the entire integral from $\tau=0$ to $\tau=\infty$ results in the green dashed and dotted lines. 


\section{Dielectric response computation: Two partial examples}
\subsection{Maxwellian plasmas without drifts}

 \textcolor{black}{Performing the integral over the parallel velocity for a Maxwellian distribution function}
\begin{equation} 
F_{o,Max}=exp[-v^2/[2v_t^2]]/[2\pi v_{th}^2]^{3/2}
\end{equation}
(where $\int F_{o,Max}=1$) \textcolor{black}{ yields}  the Fried-Conte function $\mathcal{Z}(\zeta)$ or moments of it: \textcolor{black}{The Kennel-Engelman operator yields 3 flavours of terms, namely 
\begin{itemize}
\item (i) terms proportional to $v_{//}^2$ in the numerator, important for the electron Landau damping term in the dielectric tensor element $K_{//,//}$, 
\item (ii) terms proportional to $v_{//}$ (coupling the parallel and perpendicular components of the electric field) and - finally - 
\item (iii) terms not having $v_{//}$ in the numerator at all (purely perpendicular dynamics). 
\end{itemize}} 
Isolating the parallel dynamics while using the simplified expression obtained we indeed get integrals of the form
$$\int dk_{//} \int dk'_{//} exp[i(k'_{//}-k_{//})x_{//}] I_M F^*_{\vec{k},\alpha} E_{\vec{k}',\beta}$$
\begin{equation} 
=\int dk_{//} \int dk'_{//} exp[i(k'_{//}-k_{//})x_{//}] \Bigg[ \int dv_{//}  \frac{v_{//}^M}{i[N\Omega + \mathtt{k}_{//}v_{//}-\omega]} exp \Big [-v_{//}^2/[2v_{th}^2 ] \Big ] \Bigg ] F^*_{\vec{k},\alpha} E_{\vec{k}',\beta}
\end{equation}
The 3 relevant innermost integrals are
$$I_0=\frac{\pi^{1/2}}{i\mathtt{k}_{//}} \mathcal{Z}(\zeta)$$
$$I_1=\frac{[2\pi]^{1/2}v_{th}}{i\mathtt{k}_{//}} [1+ \zeta \mathcal{Z}(\zeta)]$$
\begin{equation} 
I_2=\frac{2\pi^{1/2}v_{th}^2}{i\mathtt{k}_{//}} \zeta [1+ \zeta \mathcal{Z}(\zeta)]
\end{equation}
in which $\zeta=[\omega-N\Omega]/[\mathtt{k}_{//}2^{1/2}v_{th}]$. When $\mathtt{k}_{//}\rightarrow 0$ (so $\zeta \rightarrow \infty$) the $v_{//}$ dependence disappears from the velocity integrals and the integrals become simple moments. Taking the asymptotic expression for the Fried-Conte function 
\begin{equation} 
\mathcal{Z}(\zeta)=-\frac{1}{\zeta} [1+\frac{1}{2\zeta^2}+\frac{3}{4\zeta^4}]
\end{equation}
allows to find the proper expression for that limit:
$$I_0\approx - \frac{\pi^{1/2}}{i\mathtt{k}_{//}\zeta} \approx  \frac{i[2\pi]^{1/2}v_{th}}{[\omega-N\Omega]}$$
$$I_1\approx-\frac{[2\pi]^{1/2}v_{th}}{i\mathtt{k}_{//}2\zeta^2} \approx \frac{i[2\pi]^{1/2}v_{th}^3}{[\omega-N\Omega]^2}\mathtt{k}_{//}  \rightarrow 0$$
\begin{equation} 
I_2\approx-\frac{2\pi^{1/2}v_{th}^2}{i\mathtt{k}_{//}2\zeta}\approx \frac{i[2\pi]^{1/2}v_{th}^3}{[\omega-N\Omega]}
\end{equation}
These terms seem to vanish when the temperatures goes to zero but remember each direction brings an extra factor $1/(2\pi v_{th})^{1/2}$  for each of the 3 velocity directions, ensuring the integral of the distribution is 1. So $I_1$ vanishes in that limit but the other 2 integrals stay finite. Exactly at the resonance the remaining expressions seem to diverge. A sharp but finite contribution emerges instead of the resonance when accounting for collisions by replacing $\omega$ by $\omega+i\nu$ where $\nu<<\omega $ is the collision frequency. This procedure ($Im(\omega)\rightarrow 0^+$) is already implicitly used to define the collisionless damping  behaviour of the Fried-Conte function.

More general expressions can directly be found using Eq.\ref{Eq1}. When wanting to include more detail the earlier introduced expressions
\textcolor{blue}{
}
\begin{equation} 
\frac{1}{\Phi_1} \rightarrow \frac{i\pi^{1/2}}{2\alpha} Erfc(\beta) exp[\beta^2]
\end{equation}
can be used to generalise. One gets 
$$\vec{F}^*.\overline{\overline{K}}.\vec{E}=[1-\sum_\alpha \frac{\omega^2_\alpha}{\omega^2}]\vec{F}^*_{\vec{k}'}.\overline{\overline{1}}.\vec{E}_{\vec{k}} -2\pi \sum_\alpha \frac{\omega^2_\alpha}{\omega^2} \sum_{N=-\infty}^{+\infty}  \int_{-\infty}^{+\infty}d\vec{k}'  \int_{-\infty}^{+\infty}d\vec{k}  \int_0^{+\infty} dv_\perp \int_{-\infty}^{+\infty}dv_{//}$$
\begin{equation}
 \Big [ N\Omega_\alpha \partial F_{o,\alpha}/\partial v_\perp+k_{//}v_\perp N\Omega_\alpha \partial F_{o,\alpha}/\partial v_{//} \Big ] L(\vec{F}^*_{\vec{k}'})L(\vec{E}_{\vec{k}})\frac{i\pi^{1/2}}{2\alpha} Erfc(\beta) exp[\beta^2]
 \label{Eq2}
 \end{equation}
The power balance corresponding to the wave equation merely requires to substitute the test function vector for the electric field in the weak form of the wave equation. Only the active power (i.e. the resonant interaction) is needed. To obtain the quasilinear diffusion operator, $\vec{F}$ is replaced by $\vec{E}$ but the velocity space integrals are performed piecewise in constants of motion space; similar to the wave equation, the Fokker-Planck equation will be solved adopting a variational form of the equation.

\subsection{Arbitrary distributions plasmas without drifts}

The wave and Fokker-Planck equations will be solved using the finite element method so the electric field and the distribution functions will be expressed in terms of local low order polynomials known on a (sufficiently refined) grid. In the resulting finite elements the necessary integrals can be computed by hand. Adopting the simplest possible model to find the dielectric response yields the resonant denominator $N\Omega +\tilde{k}_{//}v_{//}-\omega$ and hence integrating across the resonance will locally yield a logarithmic contribution (see e.g. \cite{KochPLA,DVEarbitraryFo}). In Ichimaru's expression, the relevant term to account for the presence of RF fields in the wave and Fokker-Planck equations of a specie of type $\alpha$  is of the form
$$ \int_0^{+\infty} dv_\perp \int_{-\infty}^{+\infty}dv_{//} v_\perp 
$$
\begin{equation}
 \Bigg [  -2\pi \frac{\omega^2_\alpha}{\omega^2} \sum_{N=-\infty}^{+\infty} 
 \int_{-\infty}^{+\infty}d\vec{k}'  \int_{-\infty}^{+\infty}d\vec{k}  
 \frac{N\Omega_\alpha / v_\perp \partial /\partial v_\perp+k_{//} \partial /\partial v_{//}}{N\Omega +\tilde{k}_{//}v_{//}-\omega}L(\vec{E}^*_{\vec{k}'})L(\vec{E}_{\vec{k}}) \Bigg ] F_o
 \label{Eqextra}
 \end{equation} 
 (where the first $\vec{E}_{\vec{k}'}$ needs to be substituted for $\vec{F}_{\vec{k}'}$ when writing down the wave equation) in which the integrals will be split into a sum of sub-integrals between grid points: $\int dv_{\perp} dv_{//}=\sum_{i,j}\int_{v_{\perp,i}}^{v_{\perp,i+1}} dv_\perp \int_{v_{//,j}}^{v_{//,j+1}} dv_{//}$. 
 Performing the integration of $1/[N\Omega +\tilde{k}_{//}v_{//}-\omega]$ over $v_{//}$ yields - apart from the principal contribution - a delta function contribution $i\pi/|\tilde{k}_{//}|$ picked up at the \textcolor{black}{resonance}. 
 Indeed, the only contribution giving rise to net energy transfer between the wave and the particles is picked up when $N\Omega+\tilde{k}_{//}v_{//}-\omega=0$ i.e. the RF term required in the Fokker-Planck equation is the sum of all values at the resonance crossings. Defining $v_{//,res}=[\omega-N\Omega]/\tilde{k}_{//}$ and adding a very small imaginary (collision or decorrelation frequency) correction to the driver frequency to ensure causality, we can evaluate the integral close to the resonant parallel velocity: 
 
\begin{figure}   
 \begin{center}
 \includegraphics[width=3.1in]{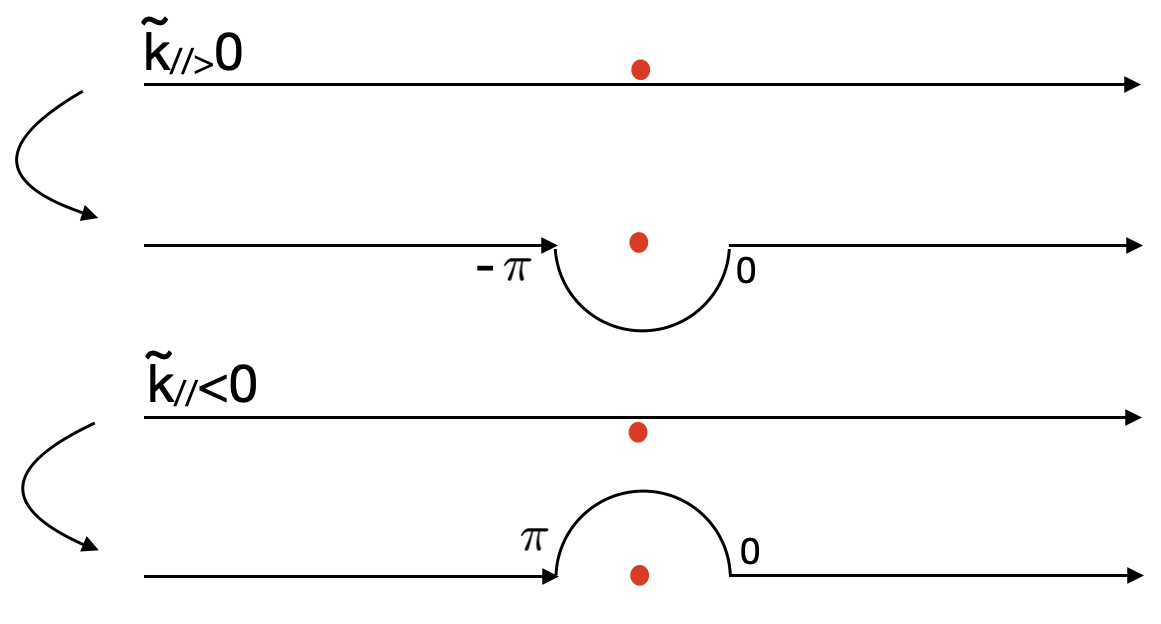}
\caption{Deformed contour lines for both signs of $\tilde{k}_{//}$.} 
 \end{center}  
  \label{fig:contouri}
\end{figure}

 $$\int dv_{//} \frac{\mathcal{H}}{[N\Omega +\tilde{k}_{//}v_{//}-\omega]}\rightarrow   \int dv_{//} \frac{\mathcal{H}}{[N\Omega +\tilde{k}_{//}v_{//}-\omega-i\nu]}$$
 \begin{equation}
= \frac{1}{\tilde{k}_{//}}\int dv_{//} \frac{\mathcal{H}}{[v_{//}-v_{//,res}-i\frac{\nu}{\tilde{k}_{//}}]}= \frac{1}{\tilde{k}_{//}}\int d\zeta \frac{\mathcal{H}}{[\zeta-i\frac{\nu}{\tilde{k}_{//}}]}
 \end{equation}
A non\textcolor{black}{-}zero imaginary part represents damping or excitation; to isolate it we take the real part of $-i$ times the expression. That can only happen close to the pole in the integrand. The pole lies just \textcolor{black}{above} the real axis when $\tilde{k}_{//}>0$ and just \textcolor{black}{under}  it when $\tilde{k}_{//}<0$. 
Changing the integration variable $\zeta \rightarrow |\zeta| e^{[ i \alpha ]} $ where $|\zeta|$ is very small but finite and positive, yields the result $i \mathcal{H}_{res} [0-(-\pi) ] / \tilde{k}_{//} $ for positive and $i\mathcal{H}_{res} [ 0 - \pi  ] / \tilde{k}_{//} $ for negative $\tilde{k}_{//}$. 
Both results can be written as $i \pi \mathcal{H}_{res} / | \tilde{k}_{//} | $.

Note that for energy to be transferred from the waves to the particles $[N\Omega_\alpha / v_\perp \partial /\partial v_\perp+k_{//} \partial /\partial v_{//}]$ needs to be negative. Note also that when $\tilde{k}_{//}$ is zero, there is no $v_{//}$ dependence in the denominator and hence there is no resonant denominator along the integration path. Consequently, there is no damping at all. Likewise, when $v_{//,res}$ does not lie between $v_{//,j}$ and $v_{//,j+1}$ there is no damping either since \textcolor{black}{the} phase has the same sign at both ends of the integration interval.

\begin{figure}   
 \begin{center}
 \includegraphics[width=5.5in]{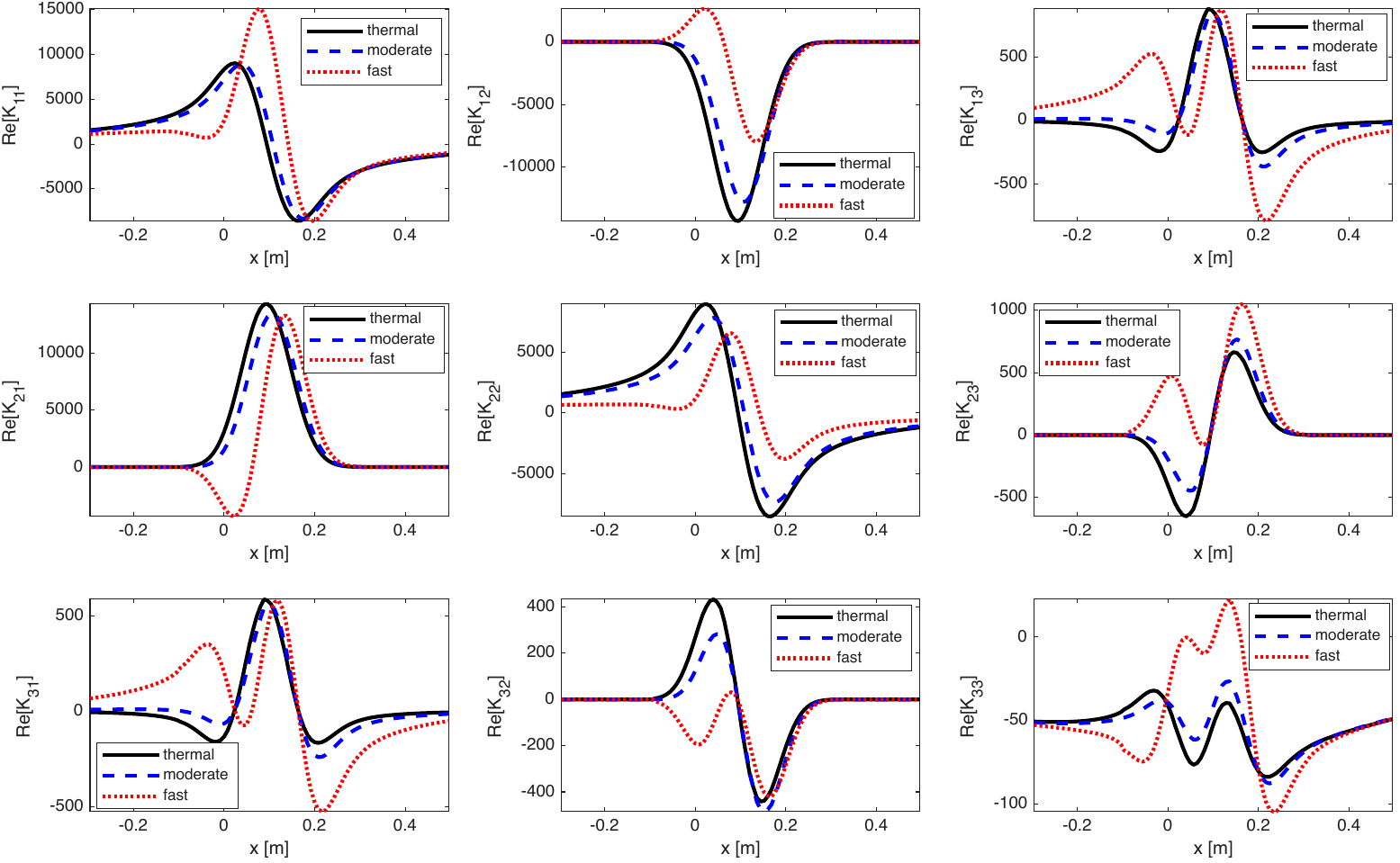}
 \includegraphics[width=5.5in]{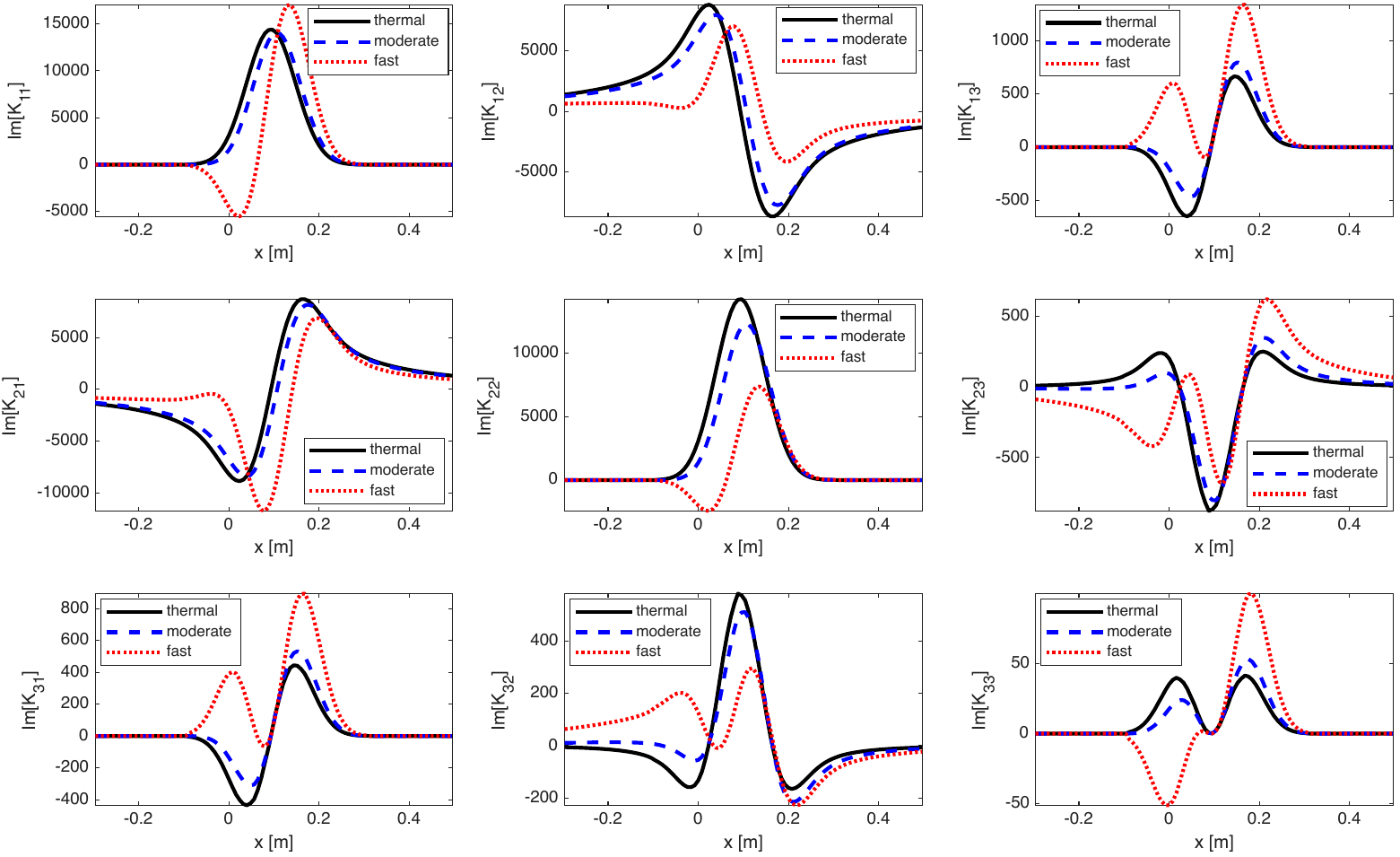}
\caption{\textcolor{black}{Dielectric tensor partial building block for $(k_{\perp},k_{//},k'_{\perp},k'_{//})=(30,7,20,4)/m$ when studying 3 types of  H minority with the same temperature in the parallel direction ($T_{0,//}=7keV$; only the contribution from H is plotted): (i) the full black line is thermal H with temperature $T_{0,\perp}=7keV$ in the centre, (ii) the dashed blue line represents a moderate tail with $T_{0,\perp}=112keV$ and (iii) the dotted red line represents a fast tail with $T_{0,\perp}=700keV$. The top 9 subplots depict the real part of the dielectric response. The bottom 9 show the corresponding imaginary parts. }}
 \label{fig:all3lH} 
 \end{center}  
\end{figure}

As an illustration, the dielectric response for a bi-Maxwellian is computed for a specific $\vec{k}$ and $\vec{k}'$. The total response is to be assembled by summing over (a discrete subset of) physically relevant modes of the spectra of $\vec{E}$ and $\vec{F}$. Figure \ref{fig:all3lH}  
shows the dielectric response element contributions from a Hydrogen minority assuming $\psi=tan^{-1}[k_{\perp,2}/k_{\perp,1}]=0$ and $(k_{\perp},k_{//},k'_{\perp},k'_{//})=(30,7,20,4)/m$. The contributions from the other plasma constituents have not been included. Typical reference JET $(H)-D$ parameters were adopted: 
central electron density $7\times 10^{19}/m^3$, $X[H]=\mathcal{N}_H/\mathcal{N}_e=5\%$. The frequency is chosen to have H fundamental cyclotron resonance heating in the centre: $f=51MHz$, $B_o=3.45T$; this places the cold cyclotron layer at $x\approx 0.1m$. The tensor elements are depicted for a series of $x=R-R_o$ in the neighbourhood of the resonance. The all-FLR dielectric tensor series has been limited to $|N_{harm}|\le 3$. The figure depicts all 9 elements of the partial dielectric "building block" for 3 types of Hydrogen minorities with the same parallel temperature of $T_{//}=7keV$: A reference thermal population with the same perpendicular temperature of $T_\perp=7keV$ in the perpendicular dimension, a moderate tail with $T_\perp=16T_{//}$, and a fast tail with $T_\perp=100T_{//}$. The tensor elements of the tail of moderate temperature crudely overlap those of the thermal H minority but significant differences are observed when the H tail is highly anisotropic. Note that the imaginary part locally becomes negative. This is a consequence of the large anisotropy in velocity space, yielding what Stix labeled as "cyclotron overstability" when $\omega T_\perp -N\Omega(T_{\perp}-T_{//})<0$ \textcolor{black}{\cite{Stix}} allowing wave excitation instead of wave damping to occur.  

\section{Brief summary and discussion}

The present paper is devoted to sketching a framework to allow a fully selfconsistent treatment of the wave and Fokker-Planck equations allowing to retain finite Larmor radius corrections \textcolor{black}{as} well as parallel and perpendicular gradients with respect to the magnetic field. It is inspired by recent encouraging results \textcolor{black}{\cite{DVE_Shelf_RF,DVE_quasimodes}} to solve the wave equation by making use of well-known classical results while casting the problem \textcolor{black}{in} a form identical to what traditional finite element solving would yield, the only difference being the way the needed coefficients of the equivalent linear system are assembled when formulating the equations in variational form. The work is to be seen as a complement to other recent efforts seeking to push the analytical expressions further while the present work leaves the burden of computational effort solely on \textcolor{black}{a sufficiently powerful} computer while exploiting well known expressions describing the wave-particle interaction. The philosophy is to construct elementary building blocks needed in both the wave and Fokker-Planck equation. These are first given in the traditionally most used limit where the guiding center motion is simply along the magnetic field. Then the expressions are generalised to the case where drifts of the guiding centre away from magnetic surfaces are incorporated. A few illustrations are provided for key points of interest. Equally, an example of the computation of the dielectric response is provided for a non-Maxwellian distribution. 

\textcolor{black}{A key question - unanswered at the moment - is whether the computations to assemble the "building blocks" can be done fast enough to be practical. Aiming at solving the integro-differential wave equation while being able to exploit powerful finite element tools presently existing, recent efforts allow to reduce the computational effort by making use of justified simplifications (e.g. only retaining retaining dominant couplings  \cite{DVEfastAORSA}, ensuring the inverse Fourier transform can be done by hand  \cite{Bude} or introducing a Fourier representation of the finite element base functions to avoid needing to integrate global mode base functions $exp[i\vec{k}.\vec{x}]$ over the entire physical domain \cite{DVE_quasimodes}). It remains useful, though, to keep looking into ways to keep the computational time needed as small as possible while ensuring the key physics effects are accounted for. Avoiding directly using the Fourier representation is one of the explored roads. Recent and partly still ongoing efforts by e.g. Machielsen and Lamalle \cite{Machielsen,Lamalle_kernel_0,Lamalle_kernel} go in that direction. In view of the fact that much of the computational effort results from the expansion $exp[i b sin \alpha]=\sum_{-N=\infty}^{+\infty}J_N(b)exp[iN\alpha]$ - which is intuitive as it allows to separate the effect of various cyclotron harmonics but requires evaluation of a series of Bessel functions - is to abandon making that expansion while seeking to integrate the original phase factor $exp[i\Phi]$ directly using the "snapshot" approach of locally adopting a truncated Taylor series expansion. Something partly along this line was already proposed by Qin \cite{Qin} who replaced the development of the susceptibility tensor in terms of infinite sums of products of Bessel functions by double integrals over the gyroperiod, yielding a compact expression for the dielectric tensor in which the infinity of cyclotron harmonics appears via the denominator $sin(\pi [\omega-k_{//}v_{//}]/\Omega)$. }

\end{document}